\documentclass[twocolumn,apj,numberedappendix]{emulateapj}
\shorttitle{Fringe-Rate Filtering}
\shortauthors{Parsons, et al.}

\usepackage{amsmath}
\usepackage{graphicx}
\usepackage[figuresright]{rotating}
\usepackage{natbib}
\usepackage{ctable}
\newcommand{\vis}{\mathbf{v}}
\newcommand{\x}{\mathbf{x}}
\newcommand{\xhat}{\hat{\mathbf{x}}}
\newcommand{\A}{\mathbf{A}}
\newcommand{\N}{\mathbf{N}}
\newcommand{\rhat}{\hat{\mathbf{r}}}

\begin{document}
\title{Optimized Beam Sculpting with Generalized Fringe-Rate Filters}

\author{
Aaron R. Parsons\altaffilmark{1,2},
Adrian Liu\altaffilmark{1,3},
Zaki S. Ali\altaffilmark{1},
Carina Cheng\altaffilmark{1}
}

\altaffiltext{1}{Astronomy Dept., U. California, Berkeley, CA}
\altaffiltext{2}{Radio Astronomy Lab., U. California, Berkeley, CA}
\altaffiltext{3}{Berkeley Center for Cosmological Physics, Berkeley, CA}

\begin{abstract}
We generalize the technique of fringe-rate filtering, whereby visibilities 
measured by a radio interferometer are re-weighted according to their temporal variation.
As the Earth rotates, radio sources traverse through an interferometer's fringe pattern
at rates that depend on their position on the sky.  Capitalizing on this geometric interpretation
of fringe rates, we employ time-domain convolution kernels to
enact fringe-rate filters that sculpt the effective primary beam of antennas in an interferometer.   As we show,
beam sculpting through fringe-rate filtering 
can be used to optimize measurements for a variety of applications,
including mapmaking, minimizing polarization leakage, suppressing instrumental systematics,
and enhancing the sensitivity of power-spectrum measurements. We show that fringe-rate filtering arises naturally in minimum variance
treatments of many of these problems, enabling optimal visibility-based approaches to analyses of
interferometric data that avoid systematics potentially introduced by traditional approaches such as imaging. Our techniques have
recently been demonstrated in \citet{ali_et_al2015}, where new upper limits
were placed on the $21\,\textrm{cm}$ power spectrum from reionization, showcasing the ability
of fringe-rate filtering to successfully boost sensitivity and reduce the impact of systematics in
deep observations.
\end{abstract}


\section{Introduction}

In recent years, low-frequency radio interferometers have undergone dramatic changes in design.
These transformations have been driven by new science applications such as $21\,\textrm{cm}$
cosmology, where one uses the highly redshifted emission from the $21\,\textrm{cm}$ hyperfine transition
of neutral hydrogen to map our early Universe. Observers in $21\,\textrm{cm}$ cosmology seek to
measure small fluctuations (both spatially and spectrally) in a dim, diffuse background that is obscured by bright
foreground emission orders of magnitude brighter in brightness temperature. This stands
in contrast to many traditional observations in radio astronomy, which more usually target bright,
compact objects in front of a dim background, often over a small selection of frequencies. These differences have led to the design, construction, 
and usage of new interferometers that only have moderate angular resolution, but are comprised of a
large number of receiving elements with wide fields of view operating over a wide bandwidth. Examples of new interferometers that broadly fit some or all of this description include the Donald C. Backer Precision Array for Probing the Epoch of Reionization (PAPER; \citealt{parsons_et_al2010}), the Murchison Widefield Array (MWA; \citealt{tingay_et_al2013,bowman_et_al2012}), the LOw Frequency ARray (LOFAR; \citealt{van_haarlem_et_al2013}), the Canadian Hydrogen Intensity Mapping Experiment (CHIME; \citealt{bandura_et_al2014}), the MIT Epoch of Reionization experiment (MITEoR; \citealt{zheng_et_al2014}), the Large Aperture Experiment to Detect the 
Dark Ages (LEDA; \citealt{greenhill_et_al2012}), and the Hydrogen Epoch of Reionization
Array (HERA; \citealt{pober_et_al2014}). Further deviating from conventional array
designs, the PAPER, MITEoR, CHIME, HERA projects have also maximized sensitivity by choosing to place their antenna elements in regular, redundant grids \citep{parsons_et_al2012a}.

With new interferometer designs, it is natural to expect new approaches to data analysis. In this paper,
we critically examine methods for time integration. Integrating in time is a crucial step for the
high-sensitivity applications of modern low-frequency radio astronomy. Consider the measurement of
the high-redshift $21\,\textrm{cm}$ power spectrum as an example application. At the relevant redshifts ($z\sim 6$ to $20$), theoretical models suggest that this cosmological signal will be faint --- on the order of $1\,\textrm{mK}$ in brightness temperature. The noise power spectrum on such
measurements reaches comparable magnitudes only after long integration ($\gtrsim 1000\,\textrm{hrs}$)
on instruments optimized for such a measurement \citep{harker_et_al2010,parsons_et_al2012a,beardsley_et_al2013,pober_et_al2014}, and even then, often only for the largest spatial modes (depending on the instrument).
Long time-integrations are therefore crucial not only for generating the requisite sensitivity for a
detection of the cosmological signal, but also to allow faint systematics to be detected and excised
from the data.

In this paper, we extend ideas introduced in \citet{parsons_backer2009} (as well as similar ideas in \citealt{roshi_and_perley2003}, \citealt{offringa_et_al2012} and \citealt{shaw_et_al2013}) to optimize the process of combining time-ordered data. The key realization is that fringe-rate---the Fourier dual to time---is a
more natural space to enact time-averaging. Traditional time-averaging (say, a running box-car average)
is equivalent to multiplying by a sinc filter in the fringe-rate domain. Generalizing this process, the
convolution theorem ensures that time integration can be achieved by 
weighting the data in the fringe-rate domain. The fringe-rate domain provides a natural basis for time-averaging interferometric
data because astronomical sources are locked to the celestial sphere, and therefore appear at
predictable fringe-rates in the data. In particular, for a given interferometric baseline, there exists
a maximum allowable fringe-rate, beyond which there is only instrumental noise. Fringe-rate filtering
allows the clean elimination of such noise-like modes.

We place a particular emphasis on the geometric interpretation of fringe-rate filtering,
where weightings in the fringe-rate domain result in changes to an interferometer's spatial response,
effectively allowing different portions of the sky to be selected by carefully chosen fringe-rate filters.
These filters can be optimized for a number of different applications, including the measurement
of cosmological power spectra, the reduction of polarization leakage, and the downweighting of
contaminating sources far from the central regions of the sky that one is attempting to observe.
Importantly, these filters can be implemented on a per-baseline basis, providing a different view
of systematics in the data, which are often easier to identify when described baseline-by-baseline,
instead of being mixed together in an image-domain map. However, we will also show that 
optimally weighted mapmaking can also be more conveniently conceptualized in a mathematical
framework operating in the fringe-rate basis.

The rest of the paper is organized as follows. In Section \ref{sec:overview} we provide a general
overview of fringe-rate filtering, establishing an essential geometric intuition for the process. The
specific implementation that we use for the simulations in this paper are described in Section
\ref{sec:Implementation}. Section \ref{sec:bmsculpt} describes how fringe-rate filtering can be
optimized for various applications. We pay specific attention to the problem of mapmaking in
Section \ref{sec:Mapmaking}, and show that fringe-rate filtering arises naturally in that context as well.
We summarize our conclusions in Section \ref{sec:conclusion}.

\section{Overview of principle of fringe-rate filtering}
\label{sec:overview}
Generally, the interferometric response $V$ at frequency $\nu$ for two antennas in a radio interferometer is described
by the visibility function\footnote{In this section, we omit the instrumental noise contribution to the measured visibilities in order to avoid notational clutter.}
\begin{equation}
\label{eq:originalVis}
V_{b\nu}(t)=\int d\Omega \, {I_\nu(\rhat) A_\nu(\rhat,t) \exp \left[-i2\pi \frac{\nu}{c}  \mathbf{b}(t) \cdot \rhat\right]},
\end{equation}
where $I_\nu$ is the specific intensity of the sky in the direction $\rhat$,
$A_\nu$ is the geometric mean of the primary beam power patterns of the constituent antennas (henceforth known as ``the primary beam"), $\mathbf{ b}(t)$ is the baseline vector separating the two antennas in question (which is time-dependent since the baselines rotate with the Earth), and $\nu$ is the spectral
frequency.
We adopt the convention that our coordinate system is fixed to the celestial sphere, because it will be convenient for our algebraic manipulations later. However, it is equally valid to understand the time-variation of the visibilities as arising from the movement of astronomical sources through the primary beam and the fringes arising from a baseline, which are fixed to a topocentric coordinate system. For drift-scan telescopes like PAPER, CHIME, or HERA, this view is particularly powerful because then the primary beam and the fringe pattern are locked to one another, and may together be considered an enveloped fringe pattern that gives rise
to time-variation in $V_{b\nu}(t)$ as the Earth rotates.

\begin{figure}\centering
\includegraphics[width=.9\columnwidth]{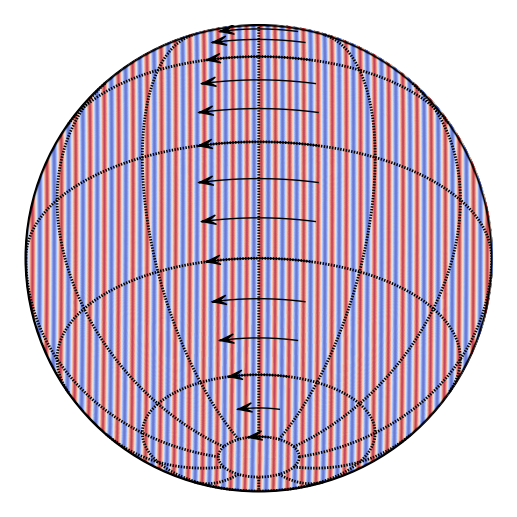}
\caption{
The fringe pattern at 150 MHz of a fiducial 30-m east-west baseline, overlaid with arrows indicating
the distance traversed by sources at various declinations over a two-hour time span centered at transit.
In a fixed time interval, sources near declination $\delta=0^\circ$ traverse more 
fringe periods than sources nearer to the celestial poles. This gives rise to different
fringe rates that can be used to distinguish sources in a time-series measured with a single baseline.
}\label{fig:ew_fringe}
\end{figure}

The rate at which angular structure on the sky moves relative to the fringe pattern---the \emph{fringe rate}---depends on the declination and hour angle. 
%
%
As an example, Figure \ref{fig:ew_fringe} illustrates the real component of the phase
variation in the fringe pattern of a 30-m east-west baseline deployed at $-30^\circ$ latitude.
Although fringes are evenly spaced in $l\equiv\sin\theta_x$, the distance a source that is locked to the celestial
sphere travels through the fringe pattern depends on its position on the sphere. This is illustrated in Figure
\ref{fig:ew_fringe} by arrows that indicate the motion of sources at differing declinations over the course
of two hours near transit.  As the source and the fringe pattern move relative to one another,  $V_\nu(t)$ oscillates
with an amplitude that is determined by the strength of the source and the amplitude of the beam response,
and a frequency that corresponds to the number of fringe periods traversed in a given time interval.  Hence, the frequency or {\it fringe-rate} of oscillations in $V_\nu(t)$
ranges from a maximum at $\delta=0^\circ$ to zero at $\delta=-90^\circ$, and can even become negative
for emission from the far side of the celestial pole.

Let us now derive this intuition mathematically, assuming a drift-scan telescope. To sort our time-variable visibilities into different fringe-rates $f$, we take the Fourier transform of our visibility over a short interval of time centered at time $t$ to get
\begin{equation}
\label{eq:OriginalFR}
\overline{V}_{b\nu} (f,t) = \int d\Omega I_\nu(\rhat)\!\! \int dt^\prime \gamma (t^\prime - t) A_\nu(\rhat,t^\prime) e^{-i2 \pi \left [(t^\prime - t)f +\frac{ \nu}{c} \mathbf{b}_{t^\prime} \cdot \rhat  \right]},
\end{equation}
where we have introduced the notation $\mathbf{b}_t \equiv \mathbf{b}(t)$, and $\gamma$ is a tapering function for the Fourier transform in time, which we assume peaks when its argument is zero, in essence shifting the origin of our transform to time $t$. If the characteristic width of $\gamma$ is relatively short, one is effectively examining the visibility over short timescales, during which its time-dependence will likely be dominated by features on the sky moving relative to fringes, and not the movement of the primary beam through the celestial sphere. We may therefore say that for short periods of time, $A_\nu (\rhat,t^\prime) \approx A_\nu (\rhat,t)$. Additionally, we may take the time-dependence of the baselines to leading order, with
\begin{eqnarray}
\mathbf{b}_{t^\prime} &\approx& \mathbf{b}_{t} + \frac{d\mathbf{b}}{dt} \Bigg{|}_{t^\prime=t} (t^\prime -t) + \dots \nonumber \\
&=& \mathbf{b}_t- (\mathbf{b}_t \times \boldsymbol \omega_\Earth) (t^\prime -t) + \dots
\end{eqnarray}
where $\boldsymbol \omega_\Earth$ is the angular velocity vector of the Earth's rotation. In the last equality, we used the fact that the time-dependence of the baselines are not arbitrary, but instead are tied to the Earth's rotation, transforming the time derivative into a cross-product with $\boldsymbol \omega_\Earth$, as one does in the analysis of solid rotating bodies. Inserting these approximations into Equation~\eqref{eq:OriginalFR} yields
\begin{eqnarray}
\label{eq:nextFR}
\overline{V}_{b\nu} (f,t) =  \int &&d\Omega  I_\nu(\rhat) A_{\nu} (\rhat,t) e^{-i2 \pi \frac{ \nu}{c} \mathbf{b}_t \cdot \rhat }   \nonumber \\
&&\cdot\int dt^\prime \gamma (t^\prime -t)  e^{i2 \pi [ \frac{\nu}{c} (\mathbf{b}_t \times \boldsymbol \omega_\Earth) \cdot \rhat -f] (t^\prime -t)} \nonumber \\
=  \int d\Omega I_\nu(\rhat)  && \!\! A_{\nu} (\rhat,t)  e^{-i2 \pi \frac{ \nu}{c} \mathbf{b}_t \cdot \rhat } \,\widetilde{\gamma} \! \left[ \frac{\nu}{c} (\mathbf{b}_t \times \boldsymbol \omega_\Earth) \cdot \rhat -f \right], \qquad
\end{eqnarray}
where $\tilde{\gamma}$ is the inverse Fourier transform of $\gamma$. To the extent that $\gamma(t)$ can be chosen to be relatively broad without violating our approximations, $\tilde{\gamma}$ will be peaked around the point where its argument is zero. Its presence in Equation \eqref{eq:nextFR} therefore acts approximately like a delta function, selecting portions of the sky that have $\rhat$ satisfying the condition $f \approx \rhat \cdot \mathbf{b} \times \boldsymbol \omega_\Earth \nu / c $.

\begin{figure}\centering
\includegraphics[width=.9\columnwidth]{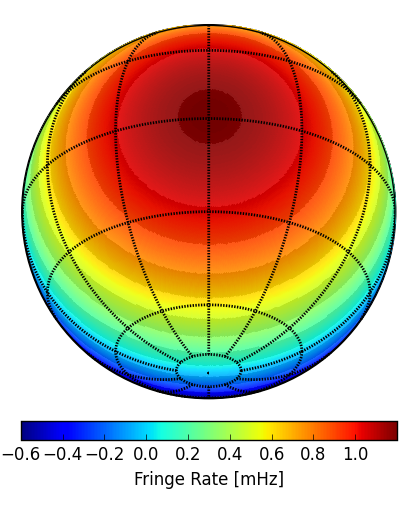}
\caption{
Fringe rate as a function of sky position, corresponding to the fringe pattern illustrated in
Figure \ref{fig:ew_fringe}.  Fringe rates peak at $1.09\,\textrm{mHz}$ at $\delta=0^\circ$, hit zero at
the south celestial pole, and become negative on the far side of the pole.  Grey shading indicates
the approximate angular regions that correspond to alternating fringe-rate bins, assuming a
fringe-rate transform taken over a two-hour time series.
}\label{fig:fringe_contours}
\end{figure}

In words, what the above derivation shows is that, as claimed, different fringe-rates correspond to different parts of the sky. This is illustrated in Figure \ref{fig:fringe_contours}, which shows shaded bands of constant fringe-rate for the same baseline as the one simulated in Figure \ref{fig:ew_fringe}. In general, contours of constant fringe-rate correspond to locations on the sky $\rhat$ that have the same degenerate combination of $\rhat \cdot \boldsymbol \omega_\Earth \times \mathbf{b} $. Note that this combination can also be rewritten as $\boldsymbol \omega_\Earth \cdot \mathbf{b} \times \rhat $ or $\mathbf{b} \cdot \rhat \times  \boldsymbol \omega_\Earth $ by cyclic permutation. Thus, if any two of $\mathbf{b}$, $\omega_\Earth$, and $\rhat$ are parallel, their cross product---and hence the fringe-rate---will be zero. For example, the fringe-rate for astronomical sources located at either celestial pole will always be zero, since $\rhat$ would then be parallel to $\boldsymbol \omega_\Earth$. Similarly, a north-south only baseline located at the equator would have $\mathbf{b}$ parallel to $\boldsymbol \omega_\Earth$, resulting in $f=0$ because in such a scenario the fringes would have no azimuthal dependence, and thus there would be no fringe-crossings as the Earth rotates relative to the sky.

Because different fringe-rates correspond to different parts of the sky, we may effectively select different portions of the sky by picking different linear combinations of fringe-rates. To see this, imagine decomposing our data into fringe-rates, and then applying a weighting function $w(f)$ before Fourier transforming back to the time domain. The result is
\begin{eqnarray}
&&V^\textrm{filt}_{b \nu}(t^\prime,t) = \int df w(f) \overline{V}_{b\nu} (f,t) e^{i 2\pi  (t^\prime-t) f} \nonumber \\
&&=  \int d\Omega I_\nu(\rhat) A_{\nu} (\rhat,t) e^{-i2 \pi \frac{ \nu}{c} \mathbf{b}_t \cdot \rhat } \nonumber \\
&& \times \int df e^{i 2\pi f (t^\prime-t)} w(f)  \widetilde{\gamma} \! \left[ \frac{\nu}{c} (\mathbf{b}_t \times \boldsymbol \omega_\Earth) \cdot \rhat -f \right].\qquad
\end{eqnarray}
Now, suppose we implement this filter in a sliding manner (a convolution) in time. That is, we repeat this process with the fringe-rate transform centered on each instant in time. With this, we become interested in only $t^\prime = t$, so the final set of filtered visibilities takes the form
\begin{equation}
\label{eq:ShrunkBeam}
V^\textrm{filt}_{b\nu}(t) = \int d\Omega \, {I_\nu(\rhat) A^\textrm{eff}_{\nu} (\rhat, t)\exp \left[-i2\pi \frac{\nu}{c}  \mathbf{b}(t) \cdot \rhat\right]},
\end{equation}
which is precisely the same as our original measurement equation, except the primary beam has been replaced by an \emph{effective primary beam}, defined as
\begin{equation}
\label{eq:EffectiveBeamDef}
A^\textrm{eff}_{\nu} (\rhat, t)\equiv A_{\nu} (\rhat, t) (w * \tilde{\gamma}) \left[ \frac{\nu}{c} (\mathbf{b}_t \times \boldsymbol \omega_\Earth) \cdot \rhat  \right],
\end{equation}
with * signifying a convolution. We thus see that by judiciously selecting fringe-rate weights, one can effectively reshape one's beam. In general, however, we cannot do so with perfect flexibility. This can be seen by once again examining the combination $\rhat \cdot \mathbf{b} \times  \boldsymbol \omega_\Earth$. For any given instant, $\mathbf{b} \times \boldsymbol \omega_\Earth$ picks out a particular direction on the celestial sphere. A ring of locations $\rhat$ on the sky at a constant angle with respect to this direction will have the same value of $\rhat \cdot  \mathbf{b} \times \boldsymbol \omega_\Earth$, and therefore the same fringe-rate. As a result, contours of constant fringe-rate always form rings on the sky, as illustrated in Figure \ref{fig:fringe_contours}. By weighting different fringe-rates, one can effectively ``turn off" (or less harshly, to simply downweight) whole contours, but never portions of a contour.

Aside from modifying the shape of one's beam, fringe-rate filtering can also be used to integrate visibilities in time. For example, if $w(f)$ is chosen in a way that suppresses high fringe rates, the effect in the time domain will be a low-pass filter that (among other features) has the rough effect of averaging together data. Enacting the time-averaging in the fringe-rate domain is particularly helpful for differentiating between noise- and signal-like modes in the time-series data. To see this, recall that the relative compactness of the $\tilde{\gamma}$ term in Equation \eqref{eq:nextFR} implies that an astronomical source located at $\rhat$ will preferentially appear at a fringe rate of $f \approx \rhat \cdot \mathbf{b} \times \boldsymbol \omega_\Earth \nu / c $ in the data. Since $\rhat \cdot \mathbf{b} \times \boldsymbol \omega_\Earth$ can never exceed $b \omega_\Earth $, the maximal fringe-rate that can be achieved by a source locked to the celestial sphere is $f_\textrm{max} = b \omega_\Earth  \nu / c$, where $\omega_\Earth \equiv | \boldsymbol \omega_\Earth|$ and $b \equiv | \mathbf{b}|$. Data at even higher fringe rates will likely be noise- rather than signal-dominated and may be filtered out safely with no loss of signal. This is a more tailored approach to reducing time-ordered data than simply averaging visibilities together in time. The latter can be viewed as a boxcar convolution in the time domain, which corresponds to applying a sinc filter in fringe-rate space. With wings that only decay as $1/f$, a sinc filter tends to incorporate data from the noise-dominated high fringe rate modes. A fringe-rate filter, in contrast, can be more carefully tailored to enhance modes that are sourced by actual emission from the celestial sphere.

In this section, we have provided some basic intuition for fringe-rate filtering, and have highlighted how it can be used for reshaping one's primary beam as well as to combine time-ordered data. In fact, these two applications are often intimately linked, since optimal prescriptions for combining time-ordered data (``mapmaking") involve re-weighting data by the primary beam \citep{T97mapmaking,Morales2009,dillon_et_al2015}. We will return to this in Section \ref{sec:Mapmaking}, where we will see that the fringe-rate filtering is a natural way to approach mapmaking in interferometric observations.

\section{Implementation}
\label{sec:Implementation}

In this section, we discuss a practical implementation of the aforementioned ideas. We will use this implementation in simulations later in the paper to illustrate various applications of fringe-rate filtering. In 
keeping with its origins as a tool for analyzing data from the PAPER array, we simulate a model array based on PAPER, deployed at a latitude of $-30^\circ$
and featuring the beam response pattern characteristic of PAPER dipole elements \citep{parsons_et_al2008,pober_et_al2012}.
The PAPER beam response pattern is illustrated in the leftmost panel of Figure \ref{fig:sim_beam}.
For these simulations, we also choose a specific baseline to examine: a pair of antennas separated by 30 m in the 
east-west direction, deployed at a latitude of $-30^\circ$, and observing at 150 MHz.  This baseline, hereafter referred to as our {\it fiducial baseline}, 
corresponds to the most repeated (and hence,
most sensitive) baseline length measured by the PAPER array in the maximum-redundancy array configuration it uses
for power spectral measurements \citep{parsons_et_al2012a,P14,ali_et_al2015}.  As such, our simulations demonstrate the performance of fringe-rate filtering in the context of the specific instrument configuration
that has recently been used to place the current best upper limits on $21\,\textrm{cm}$ emission from cosmic reionization \citep{P14,J14,ali_et_al2015}.

\begin{figure}[!t]
\centering
\includegraphics[width=\columnwidth]{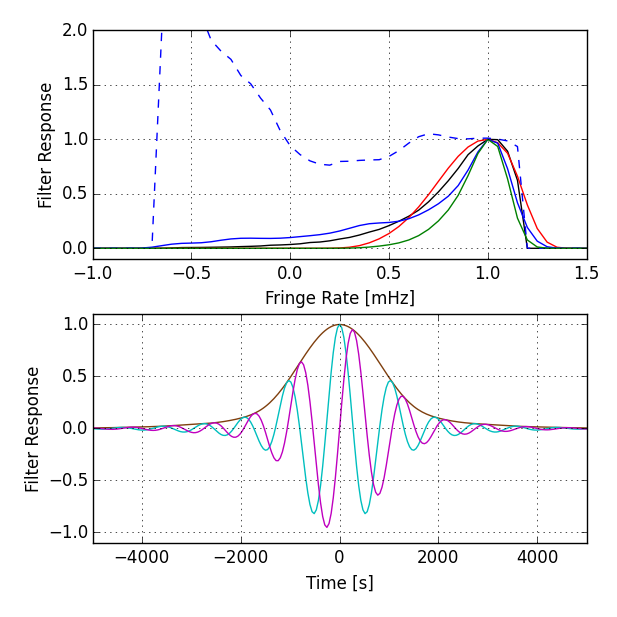}
\caption{
Top: the optimal power-spectrum sensitivity weighting in fringe-rate space 
for the XX polarization beam of our fiducial baseline (black) is overlaid with
the simple parametrization for optimal weighting (red) used in \citet{ali_et_al2015},
which excises fringe rates at risk for contamination by crosstalk, and 
applies a Blackman-Harris window in the time domain to
generate a compact time-domain convolution kernel with regions weighted below 5\% of the
peak response suppressed.
Also illustrated
are the weightings for matching the YY to the XX polarization beam (blue dashed) and
for reducing polarization leakage (blue solid) described in Section \ref{sec:polbeams}, as well
as a weighting that balances the needs of sensitivity, polarization leakage, crosstalk
removal, and off-axis foregrounds (green).  Bottom: the time-domain convolution kernel
corresponding to the red curve in the top panel.  Real and imaginary components are
illustrated in cyan and magenta, respectively, with the absolute amplitude illustrated
in brown.
}
\label{fig:fringe_rate_cut}
\end{figure}

In the following sections, we will derive a number of different fringe-rate
weights, each optimized for a different application. Often, these optimized
weights depend on the detailed properties of one's instrument, and can
therefore only be computed numerically, not analytically. For example, we will
find in Section \ref{sec:PspecOptimization} that the optimal fringe-rate
weights for power spectrum estimation involve computing the root-mean-square
(RMS) primary beam profile over contours of constant fringe rate on the sky
(such as those shown in Figure \ref{fig:fringe_contours}).  An example of the
binning of the RMS beam response in fringe rate is given by the black curve in
the top panel of Figure \ref{fig:fringe_rate_cut}.  A realistic primary beam will
frequently require empirical modeling beyond analytic forms, making it generally
difficult to derive a completely analytic expression for an optimized
fringe-rate filter profile. However, in the interest of being able to rapidly
generate filters as a function of varying baseline lengths and observing
frequencies, we frequently fit analytic forms (such as truncated Gaussians) to
the numerical profiles, as shown by the red curve in the top panel of Figure
\ref{fig:fringe_rate_cut}.  As long as the numerical profiles take the
optimized forms that we will derive in Section \ref{sec:bmsculpt}, small
deviations arising from an imperfect analytic fit are unlikely to significantly
shift the final error properties of one's measurements. With the discussion of
power spectrum measurements in Section \ref{sec:PspecOptimization}, for
example, we minimize the noise variance by varying the fringe-rate weights.
Because our analytic fits to these weights start from a local minimum in noise
variance, any deviations in the weighting profile will only induce small
second-order increases in the final error bars.

%
%

The next step in implementing the fringe-rate filter is translating the analytic filter profile in fringe-rate space into 
a time-domain kernel that can be used to convolve the simulated time series of visibilities.  In effect, we implement
the fringe-rate filter as a finite impulse response (FIR) filter.  The convolution kernel corresponding to this FIR
filter is shown in the lower panel of Figure \ref{fig:fringe_rate_cut}.  Applying the fringe-rate filter as an FIR filter in
the time domain, as opposed to directly multiplying the desired filter to Fourier-transformed visibilities, has the advantage 
that flagged or missing data can be naturally excluded from the filter by neglecting
FIR taps (coefficient multiplies) that target the missing data. The summed output of the FIR filter are then renormalized to
account for the missing samples.  Another advantage of the FIR implementation of the fringe-rate filter is the potential for
windowing the time-domain filter profile.  While time-domain windowing causes further deviations from the ideal
fringe-rate filter profile, it can be used to produce a more compact time-domain kernel.
Reducing the number of time-domain samples used in the FIR filter improves the computational efficiency of the filter, 
helps limit the number of samples potentially corrupted by spurious systematics such as radio frequency interference,
and reduces boundary effects.  In \citet{ali_et_al2015}, a 12-hour data set was filtered in this manner, with boundary
effects limited to an hour on each end.  The final analysis proceeded on 8.3 hours selected from within the region
not affected by boundary effects.

\section{Applications}
\label{sec:bmsculpt}

In Section \ref{sec:Implementation}, we discussed how a fringe-rate filter can be implemented in practice once a particular form for the filter is selected. In this section, we optimize the selection of filters (or equivalently, of fringe-rate weights) for various applications in low-frequency radio interferometry. The key to this optimization will be the insight from Section \ref{sec:overview}, namely that the effect of fringe-rate filtering can be regarded as both a time integration and a modification of the spatial response of the primary beam on a per-baseline basis. Turning this around, one can identify the optimal primary beam needed for one's observations, and then reverse engineer the set of fringe-rates needed to achieve this beam in what is essentially a ``beam sculpting" operation. For concreteness, we will focus here on $21\,\textrm{cm}$ cosmology, but many of the ideas presented here are easily translatable to other applications of radio interferometry.

\subsection{Minimizing thermal noise errors in power spectrum measurements}
\label{sec:PspecOptimization}

In \citet{parsons_et_al2012a} and \citet{P14}, it was shown that estimates of the three-dimensional power spectrum of $21\,\textrm{cm}$ brightness temperature fluctuations could be obtained from a single baseline by Fourier transforming visibility data along the frequency axis (forming a ``delay spectrum"), and then taking the absolute square of the results. Here, we will show how fringe-rate weights can be chosen to maximize the sensitivity of a single-baseline-derived power spectrum.

We begin by considering a generalization of the derivation in \citet{P14}, where it was assumed that the primary beams of all elements in the interferometer are identical. We now consider the possibility of probing the power spectrum via a cross-correlation of two baselines with different primary beams. To be clear, our eventual discussion will be based on the analysis of fringe-rate filtered visibilities from a \emph{single} baseline. However, from Section \ref{sec:overview}, we saw that to a good approximation, selecting different fringe-rates is equivalent to observing the sky with different effective beams. Thus, the cross-correlation of visibilities from two different fringe-rate bins is mathematically identical to cross-correlating two baselines with different beams. To begin, suppose that the $i$th baseline consists of antenna elements with primary beam $A_i (\rhat, \nu)$. The delay-transformed visibility takes the form
\begin{equation}
\widetilde{V}_i(\mathbf{u},\eta) = \frac{2 k_B}{\lambda^2} \int  d^2 \mathbf{u}^\prime \, d\eta^\prime \widetilde{A}_i (\mathbf{u} -\mathbf{u}^\prime, \eta-\eta^\prime) \widetilde{T}(\mathbf{u}^\prime , \eta^\prime),
\end{equation}
where $\eta$ is the Fourier dual to frequency\footnote{This equation can be derived by Fourier transforming Equation 
\eqref{eq:originalVis} along the frequency axis and re-expressing the angular integral in $uv$ coordinates assuming the flat-sky approximation. However, it also makes the crucial assumption that one can neglect the frequency-dependent nature of the mapping of baseline $\mathbf{b}$ to $\mathbf{u}$ coordinate. In practice, this is only a reasonable approximation for short baselines \citep{parsons_et_al2012b,liu_et_al2014a} such as those used for power spectrum analyses in the PAPER experiment \citep{P14,J14,ali_et_al2015}.
}, $\widetilde{A}_i $ is the Fourier transform of $A_i (\rhat, \nu)$ in both the angular and spectral directions, $\widetilde{T}$ is the brightness temperature field in Fourier space, $k_B$ is Boltzmann's constant, $\lambda$ is the central observation frequency, and $\mathbf{b} = \mathbf{u} \lambda$, with $\mathbf{u} \equiv (u,v)$. From this, we can see that two baselines with different primary beams, but located at the same location on the $uv$ plane have a delay-spectrum cross-correlation given by
\begin{eqnarray}
\label{eq:visCrossCorr}
\langle \widetilde{V}_i(\mathbf{u},\eta) \widetilde{V}_j(\mathbf{u},\eta)^*\rangle = \left( \frac{2 k_B}{\lambda^2} \right)^2 \int d^2 \mathbf{u}^\prime \, d\eta^\prime  P(\mathbf{u}^\prime , \eta^\prime) \nonumber \\
\times \widetilde{A}_i (\mathbf{u} -\mathbf{u}^\prime, \eta-\eta^\prime) \widetilde{A}_j^* (\mathbf{u} -\mathbf{u}^\prime, \eta-\eta^\prime) \nonumber \\
\approx P(\mathbf{u} , \eta) \left( \frac{2 k_B}{\lambda^2} \right)^2 \int d\Omega d\nu A_i (\rhat,\nu) A_j (\rhat,\nu), \quad
\end{eqnarray}
where angular brackets $\langle \dots \rangle$ denote an ensemble average over possible realizations of a random temperature field. In the first equality, we assumed that this field is a translation-invariant Gaussian random field specified by a power spectrum $P(\mathbf{u}, \eta)$, so that
\begin{equation}
\langle \widetilde{T}(\mathbf{u} , \eta) \widetilde{T}^*(\mathbf{u}^\prime , \eta^\prime)\rangle = \delta^{D} (\mathbf{u} - \mathbf{u}^\prime) \delta^D (\eta - \eta^\prime) P(\mathbf{u}, \eta).
\end{equation}
In the second equality, we made the approximation that for reasonably broad primary beams, $\widetilde{A}_i $ and $\widetilde{A}_j$ tend to be rather localized, which allows the comparatively broader $P(\mathbf{u}, \eta)$ to be factored out of the integral.\footnote{Although see Section IV A of \citet{liu_et_al2014b} for some limitations of this approximation.} Following this, we used Parseval's theorem to rewrite the integral over $(\mathbf{u},\eta)$ space as an integral over solid angle and frequency.

Rearranging Equation \eqref{eq:visCrossCorr} gives an expression for the true power spectrum in terms of the cross-correlation function of two delay-space visibilities. With real data, however, one cannot perform the ensemble average on the left-hand side of Equation \eqref{eq:visCrossCorr}. Omitting this ensemble average, the copy of the power spectrum on the right-hand side becomes an \emph{estimator} $\widehat{P}$ of the true power spectrum $P$. Introducing the definition
\begin{equation}
\label{eq:Omega_ij_def}
\Omega_{ij}^{pp} \equiv \frac{1}{B} \int d\Omega d\nu A_i (\rhat,\nu) A_j (\rhat,\nu),
\end{equation}
where $B$ is the bandwidth over which observations are made, our estimator takes the form
\begin{equation}
\label{eq:pspecEstCrossCorr}
\widehat{P} (\mathbf{k}) = \left( \frac{\lambda^2}{2 k_B} \right)^2 \frac{X^2 Y}{\Omega^{pp}_{ij} B} \widetilde{V}_i(\mathbf{u},\eta) \widetilde{V}_j(\mathbf{u},\eta)^*,
\end{equation}
where we have written the power spectrum in terms of cosmological Fourier coordinates $\mathbf{k}$, which are related to the interferometric Fourier coordinates by $(X k_x, X k_y, Y k_z) \equiv 2 \pi (u , v, \eta)$, picking up an extra factor of $X^2 Y$ in the process,\footnote{See, e.g., \citet{liu_et_al2014a} for a detailed derivation.} with
\begin{equation}
X \equiv \frac{c}{H_0} \int_0^z \frac{dz^\prime}{E(z^\prime)}; \,\, E(z) \equiv \sqrt{\Omega_m (1+z)^3 + \Omega_\Lambda}.
\end{equation}
where $c$ is the speed of light, $z$ is the redshift of observation, $H_0$ is the Hubble parameter, $\Omega_m$ is the normalized matter density, $\Omega_\Lambda$ is the normalized dark energy density, and
\begin{equation}
Y \equiv  \frac{c(1+z)^2}{\nu_{21} H_0 E(z)},
\end{equation}
where $\nu_{21} \approx 1420\,\textrm{MHz}$ is the rest frequency of the $21\,\textrm{cm}$ line. In the special case where there is just a single primary beam, we may set $i=j$ and drop the subscripts for brevity, and Equation \eqref{eq:pspecEstCrossCorr} reduces to
\begin{equation}
\label{eq:P14est}
\widehat{P} (\mathbf{k}) = \left( \frac{\lambda^2}{2 k_B} \right)^2 \frac{X^2 Y}{\Omega_{pp} B} | \widetilde{V}(\mathbf{u},\eta) |^2,
\end{equation}
where
\begin{equation}
\Omega_{pp} \equiv \frac{1}{B} \int d\Omega d\nu |A (\rhat,\nu)|^2,
\end{equation}
which is the relation found in \citet{P14}.

Having established these results, let us re-interpret Equation \eqref{eq:pspecEstCrossCorr} as an estimator for the power spectrum from the cross-multiplication of two different discretized fringe rate bins (as opposed to the cross-multiplication of baselines with different primary beams). We are free to re-interpret our estimator in this way because of the discussion in Section \ref{sec:overview}, where we showed that each visibility could be thought of as being comprised of different fringe rate contributions, each probing a different ring on the celestial sphere. Each fringe-rate therefore has its own effective primary beam, enabling our re-interpretation. That Equation \eqref{eq:pspecEstCrossCorr} involves the cross-multiplication of visibilities after they have been delay-transformed over the frequency axis is not a problem, since the Fourier transforms required to enact the delay transform and the fringe-rate transform commute with one another.

Equation \eqref{eq:pspecEstCrossCorr} allows a power spectrum to be estimated from the cross-multiplication of any pair of fringe-rate bins. To increase signal-to-noise on the measurement, however, one ought to form all possible cross-multiplied pairs, which can then combined into a single power spectrum estimate via a weighted average. Suppressing the arguments of $\widehat{P}$ and $\widetilde{V}$ for notational cleanliness, we can write
\begin{equation}
\label{eq:wVV}
\widehat{P} = \sum_{ij} g_{ij} \widetilde{V}_i \widetilde{V}_j^*,
\end{equation}
where $g_{ij}$ is the weight assigned to the cross-multiplication of the $i$th and $j$th fringe-rate bins. Our goal is to select weights that minimize the error bars on the final power spectrum estimate. 

For our optimization exercise, assume that errors are due to instrumental thermal noise only. If the $i$th fringe-rate bin has a noise contribution of $n_i$, the noise contribution to our estimator is
\begin{equation}
\widehat{P}_\textrm{noise} = \sum_{ij} g_{ij} n_i n_j^*.
\end{equation}
The error bar corresponding to this noise contribution is given by the square root of its variance, which takes the form
\begin{eqnarray}
\label{eq:NoiseVar}
\textrm{Var}(\widehat{P}_\textrm{noise} ) &\equiv& \langle \widehat{P}_\textrm{noise}^2 \rangle - \langle \widehat{P}_\textrm{noise} \rangle^2 \nonumber \\
&= & \sum_{ijkm} g_{ij} g_{km} \left[ \langle n_i n_j^* n_k n_m^* \rangle - \langle n_i n_j^*\rangle \langle n_k n_m^* \rangle \right] \nonumber \\
&=&  \sum_{ij} g_{ij} g_{ji}\sigma^4,
\end{eqnarray}
where in the last equality we assumed that the noise is Gaussian, enabling the fourth moment term to be written as a sum of second moment (variance) terms. We further assumed that the real and imaginary components of the noise are uncorrelated with each other and between different fringe-rate bins, so that if $n_i \equiv a_i + i b_i$, we have $\langle a_i a_j \rangle = \langle b_i b_j \rangle = \delta_{ij} \sigma^2/2 $ and $\langle a_i b_j \rangle= 0$ for all $i$ and $j$.

In minimizing the noise variance, care must be taken to ensure that there is no signal loss in the power spectrum estimation. To do so, we first note that taking the ensemble average of $\widehat{P}$ gives
\begin{equation}
\langle \widehat{P} \rangle = \sum_{ij} g_{ij} \langle \widetilde{V}_i \widetilde{V}_j^* \rangle = S \sum_{ij} g_{ij} \Omega_{ij} P,
\end{equation}
where we used an ensemble-averaged version of Equation \eqref{eq:pspecEstCrossCorr} to relate the true cross-correlation to the true power spectrum, and defined $S \equiv ( B / X^2 Y) ( 2 k_B / \lambda^2 )^2$. Ensuring that there is no signal loss is thus tantamount to requiring that $ S \sum_{ij} g_{ij} \Omega_{ij} =1$, so that $\langle \widehat{P} \rangle = P$. We may impose this constraint by introducing a Lagrange multiplier $\lambda$ in our minimization of the noise variance, minimizing
\begin{equation}
\mathcal L = \sum_{ij} g_{ij} g_{ji} - \lambda \sum_{ij} w_{ij} \Omega_{ij},
\end{equation}
where both $\sigma^4$ and $S$ have been absorbed into our definition of $\lambda$. Differentiating with respect to each element and setting the result to zero gives an optimized weight given by $g_{km} \propto \Omega_{km}$, and normalizing according to our constraint yields
\begin{equation}
\label{eq:PspecOptWeights}
g_{km} = \frac{\Omega_{km}}{S \sum_{ij} \Omega_{ij}^2}.
\end{equation}

To make intuitive sense of this, let us make a few more approximations. The key quantity here is $\Omega_{ij}$, which we can see from Equation \eqref{eq:Omega_ij_def} is the overlap integral between the effective primary beams of the $i$th and $j$th fringe-rate bins. In Section \ref{sec:overview}, we saw that if one takes the fringe-rate Fourier transform over a wide enough window in time, different fringe-rates map to different portions of the sky with relatively little overlap. If this is indeed the case, $\Omega_{ij}^{pp}$ vanishes unless $i=j$. Defining
\begin{equation}
\label{mu_ij_def}
\mu_i  \equiv \frac{1}{B} \int d\Omega d\nu A_i(\rhat,\nu)^2,
\end{equation}
we have $\Omega_{ij}^{pp} \equiv \delta_{ij} \mu_i$, so our optimal estimator for the power spectrum (combining Equations \ref{eq:wVV}, \ref{eq:PspecOptWeights}, and \ref{mu_ij_def}) reduces to
\begin{equation}
\widehat{P} = \frac{1}{S \sum_j \mu_j^2}\sum_i \mu_i | \widetilde{V}_i |^2 .
\end{equation}
Suppose we now define $\mu_i^{1/2} \widetilde{V}_i$ to be an optimally weighted visibility in fringe-rate space. Transforming back to the time domain using Parseval's theorem, one obtains
\begin{equation}
\label{eq:finalEst}
\widehat{P}(\mathbf{u}, \eta) = \left( \frac{\lambda^2}{2 k_B} \right)^2 \frac{X^2 Y}{B\sum_j \mu_j^2} \sum_i |\widetilde{V}^\textrm{opt} (\mathbf{u}, \eta; t_i)|^2,
\end{equation}
where the optimally filtered visibility in the time domain $\widetilde{V}^\textrm{opt} (\mathbf{u}, \eta; t_i)$ is given by
\begin{eqnarray}
\widetilde{V}^\textrm{opt} (\mathbf{u}, \eta; t_i) &\equiv& \sum_i e^{i 2 \pi f t} \mu_i^{1/2} \widetilde{V}_i \nonumber \\
&=&  \sum_i e^{i 2 \pi f t} \left[\frac{1}{B} \int d\Omega d\nu A_i(\rhat,\nu)^2\right]^\frac{1}{2} \widetilde{V}_i. \qquad
\end{eqnarray}
This is a rather interesting result, in that the optimal power spectrum estimator for a single baseline interferometer consists of a squared statistic (i.e., one with no phase information) integrated in time. This may seem counterintuitive, particularly if one is accustomed to more conventional techniques where images are formed from the visibilities and averaged down before any squaring steps. There, it is crucial to average in time \emph{before} squaring, because data from different time steps can be sourced by the same Fourier modes on the celestial sphere. Integrating before squaring allows information from these modes to be coherently averaged together (since phase information has yet to be discarded), resulting in instrumental noise that integrates down as $1/\sqrt{t}$. This then becomes a $1/t$ dependence for the error bars on the final (squared) power spectrum results, and is a much quicker reduction of instrumental noise than if the data had been squared first, which would have resulted in a $1/\sqrt{t}$ dependence on the power spectrum errors.

In our derivation, we showed that the optimal power spectrum estimator can in fact be obtained by squaring before integrating, \emph{provided} the power spectra formed at each time instant are first fringe-rate filtered with weights $ \mu_i^{1/2}$, i.e., where each fringe-rate is weighted by the RMS primary beam within the corresponding constant-fringe-rate contour on the sky. Essentially, the pre-processing step of fringe-rate filtering (with these specific weights) replaces the independent time samples with a set of correlated visibilities that have effectively already been coherently integrated in time. Note that these weights are \emph{not} generally the same as the ones derived in Section \ref{fringeRateIntro} for optimal mapmaking, where measurements will essentially be weighted by an additional factor of the primary beam in fringe-rate space, rather than by the RMS beam weighting suggested here. Put another way, to obtain the full power spectrum sensitivity of an interferometer, it is insufficient
to simply square the Fourier amplitudes outputted from a map, even if the mapmaking was optimized to minimize the error bars of the map.  Forming the power spectrum in such a way would be equivalent to restricting $g_{ij}$ to a form separable in $i$ and $j$. This restriction precludes the form for $w_{ij}$ given by Equation \eqref{eq:PspecOptWeights}, which minimizes the error bars of the power spectrum.

Importantly, the result that we have derived here applies only when one is attempting to measure a power spectrum with a single baseline (or multiple baselines with the same baseline vector $\mathbf{b}$). This is a reasonable limit to work in for arrays such as PAPER, where a large fraction of the array's sensitivity comes from instantaneously redundant baselines \citep{parsons_et_al2012a}. For arrays that have less instantaneous redundancy, it becomes more important to combine data from multiple baselines. If multiple baselines are involved, Equation \eqref{eq:finalEst} no longer reduces to a single sum over the time axis. Said differently, it is no longer true that the full sensitivity of an array can be obtained by averaging together time-slice-by-time-slice estimates of the power spectrum estimation from fringe-rate filtered data. Instead, the optimal estimator involves a double sum over time, since with multiple baselines of roughly the same length, it is possible for baselines to rotate into one another on the $uv$ plane. That is, rotation synthesis becomes an important contribution to an interferometer's sensitivity.

We note that following fringe-rate filtering, the normalization of the power
spectrum estimator must be modified accordingly. This can be seen in Equation
\eqref{eq:finalEst}, where the scalar quantities in front of the sum are
different than those found in Equation \eqref{eq:P14est}, which is applicable
to non-fringe rate filtered data. More generally, suppose we consider an
auto-correlation-only estimator of the form
\begin{equation}
\label{eq:CountlessPhats}
\widehat{P} = \sum_{i} w_{i}^2 | \widetilde{V}_i |^2 = \sum_i | w_i \widetilde{V}_i |^2,
\end{equation}
where if $w_i$ is set to $\mu_i^{1/2}$, we recover the optimized estimator derived
above. Keeping $w_i$ arbitrary here will be useful in later sections,
when we examine estimators that are purposely non-optimal as far as
instrumental noise sensitivity is concerned, but may provide
better final results due to a better mitigation of systematics. Like before,
we may impose the condition that $\langle \widehat{P} \rangle = P$, which
requires that the weights satisfy $S \sum_i w_i^2 \mu_i = 1$ (where we have
once again invoked the approximation that $\Omega_{ij}^{pp}=0$ if $i \neq j$).
Equivalently, we may leave our weights initially unnormalized as one forms
the combination $w_i \widetilde{V}_i $ in Equation \eqref{eq:CountlessPhats},
instead compensating with a normalizing denominator. Writing the estimator
in this way, one obtains
\begin{equation}
\label{eq:DunnoWhatToCallThis}
\widehat{P} = \frac{\sum_i | w_i \widetilde{V}_i |^2}{S \sum_j w_j^2 \mu_j} =\left( \frac{\lambda^2}{2 k_B} \right)^2 \frac{X^2 Y}{ B}   \frac{\sum_i | \widetilde{V}^\textrm{filt} (t_i) |^2}{\sum_j w_j^2 \mu_j},
\end{equation}
where in the last equality we re-inserted the definition of $S$ and invoked Parseval's theorem in the numerator
to write the fringe-rate filtered visibilities in the time domain. Our expression now takes precisely the same form
as Equation \eqref{eq:P14est}, except with fringe-rate filtered visibilities $\widetilde{V}^\textrm{filt} (t)$ instead 
of the original visibilities, and $\sum_j w_j^2 \mu_j$ as a normalizing beam area instead of the integrated square
beam $\Omega_{pp}$. That the estimator can be written in such a similar form is unsurprising,
since we showed in Equation \eqref{eq:ShrunkBeam} that fringe-rate filtered visibilities are essentially
the same as the original visibilities but with modified primary beams. Indeed, the term $\sum_j w_j^2 \mu_j$
can alternatively be computed by simply evaluating the integral for $\Omega_{pp}$ but with the
effective primary beam of Equation \eqref{eq:EffectiveBeamDef} instead of the original primary beam. 
Explicitly, the relevant integral is $\frac{1}{B} \int d\Omega d\nu A^\textrm{eff}_{\nu} ( \rhat, \nu)^2$, which becomes
\begin{equation}
\frac{1}{B} \int d\Omega d\nu A_{\nu} ( \rhat, \nu)^2 \,w^2\!\left( \frac{\nu}{c} \mathbf{b}_t \times \boldsymbol \omega_\Earth \cdot \rhat  \right),
\end{equation}
where we made the approximation that $ \tilde{\gamma}$ is a reasonably compact (delta function-like) function in Equation \eqref{eq:EffectiveBeamDef}.
Suppose we now evaluate this integral by splitting the sky into rings of constant fringe-rate, i.e., regions within which
we have $ \mathbf{b}_t \times \boldsymbol \omega_\Earth \cdot \rhat$ equal to a constant. By construction, the function $w$ is
constant within each of these regions, and denoting its value in the $j$th region as $w_j$, the integral becomes
\begin{equation}
\sum_j w_j^2 \frac{1}{B}  \int d\Omega d\nu A_j ( \rhat, \nu)^2 = \sum_j w_j^2 \mu_j ,
\end{equation}
completing our proof that the normalization factor in Equation \eqref{eq:DunnoWhatToCallThis} is simply
$\Omega_{pp}$ computed with the effective primary beam.
%

\begin{figure}\centering
\includegraphics[width=0.9\columnwidth]{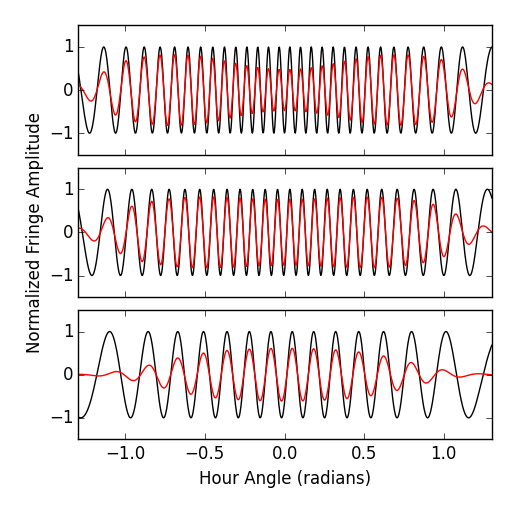}
\caption{
The real component of the fringe amplitude simulated for our fiducial 
baseline deployed at a latitude of $-30^\circ$
before (black) and after (red) the application
of the fringe-rate filter described in Section \ref{sec:Implementation}.  From top to bottom,
the panels illustrate fringes
for point sources 
passing through the fringe pattern at declinations of $0^\circ$,
$-30^\circ$, and $-60^\circ$, respectively.  In this simulation,
antenna elements have isotropic primary beams.
}\label{fig:src_track_flat}
\end{figure}

\begin{figure}\centering
\includegraphics[width=0.9\columnwidth]{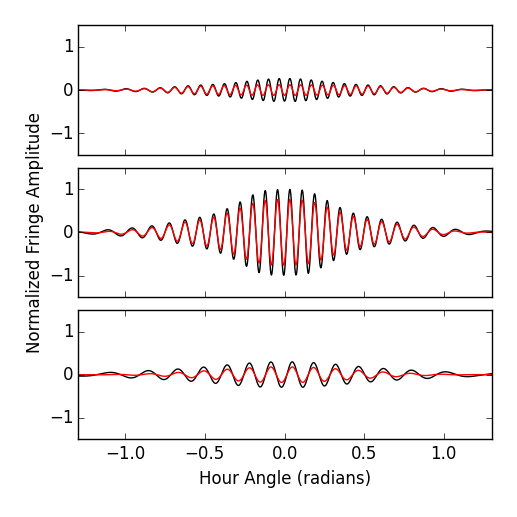}
\caption{
Same as Figure \ref{fig:src_track_flat}, but for antenna elements
with the PAPER primary beam \citep{parsons_et_al2010,pober_et_al2012}.
}\label{fig:src_track_beam}
\end{figure}

\begin{figure*}\centering
\includegraphics[width=1.9\columnwidth]{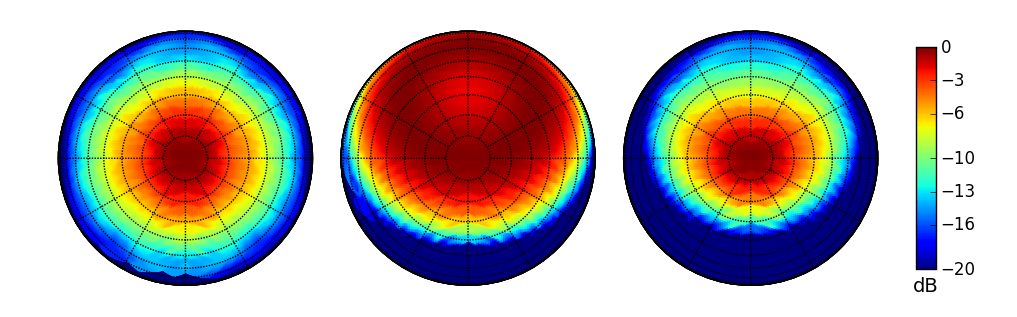}
\caption{
The effective primary beam response of a baseline, as reconstructed from point-source simulations 
described at the end of Section \ref{sec:PspecOptimization} and illustrated in Figures \ref{fig:src_track_flat} and
\ref{fig:src_track_beam}. 
The left panel shows that PAPER's model beam response is recovered from unfiltered
visibilities; the center panel illustrates the
beam weighting that results from applying a fringe-rate filter tuned to optimize sensitivity
for power spectrum measurements, assuming an isotropic primary beam; the right panel
shows the effective response after applying this fringe-rate filter to data including PAPER's model beam response.
The periodic structure along the vertical axis is an artifact of reconstructing beams from point sources 
spaced every $5^\circ$ in declination; we emphasize that they are not associated with any fundamental
structure associated with fringe-rate filtering.
}\label{fig:eff_beam}
\end{figure*}

\begin{figure*}\centering
\includegraphics[width=1.9\columnwidth]{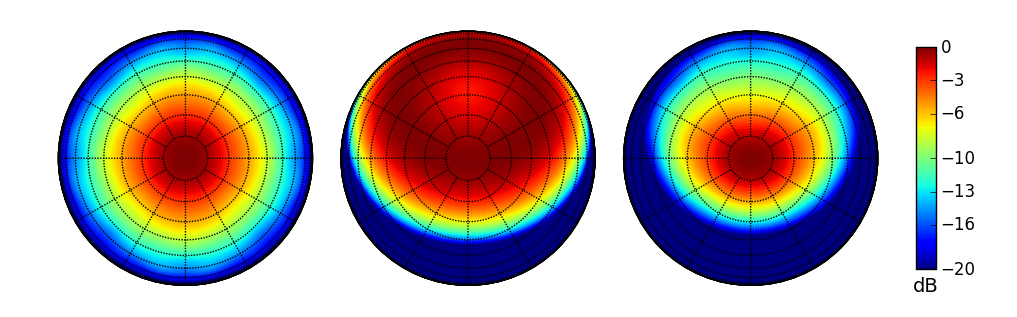}
\caption{
The model primary beam response of a baseline, simulated analytically using the interpretation
of fringe-rate filtering as a spatial filter acting along fringe-rate contours.  Panels follow
the same order as in Figure \ref{fig:eff_beam}.
}\label{fig:sim_beam}
\end{figure*}

To be conservative, however, it is important to verify our analytic results using numerical
simulations, since a number of approximations were made in our derivations. We use simulated observations 
of individual point sources positioned at declinations in $5^\circ$ increments, traversing
through the primary beam and fringe pattern as a function of time.  As
illustrated in Figure \ref{fig:src_track_flat}, we inject point sources of unity
flux density and apply the FIR implementation of fringe-rate filtering described in Section \ref{sec:Implementation}
to the visibility time-series.  The change in amplitude of the filtered visibilities at each point
along the time axis can then be 
identified with a specific position on the sky for that point source.  We bin these responses on a
HEALpix spherical grid \citep{gorski_et_al2005} to determine the effective beam response
at that location.  We perform these simulations using both an isotropic primary beam (in Figure \ref{fig:src_track_flat})
and with the PAPER primary beam (in Figure \ref{fig:src_track_beam}).  
The effective beam responses that
we recover are shown in Figure \ref{fig:eff_beam}.  The artifacts 
that appear every $5^\circ$ in declination are a result the binned beam model imperfectly 
interpolating between source tracks at different
declinations.  Aside from these artifacts, which we emphasize are associated with the reconstruction of
the beam model and not with any features exhibited along a source track, 
these results compare well to the results shown in Figure \ref{fig:sim_beam}, 
where we show the effective beam response by looking up the filter response corresponding
to the fringe rate at each sky position, i.e., by using Equation \eqref{eq:EffectiveBeamDef}.  
As these simulations illustrate, the approximations we have made in our analytic
derivation of the effect of fringe-rate filtering on the effective beam response are valid, with errors 
dominated by the effects of binning and interpolating between the 5$^\circ$ intervals in declination
between point-source simulation tracks when reconstructing an effective beam.

As a final check, we perform simulations assigning flux to pixels on the sky with a Gaussian random distribution
and simulating the visibilities measured as a function of time for our fiducial baseline.  We compute the
power spectrum amplitude for this simulated signal and a fringe-rate filtered version according to
Equation (\ref{eq:P14est}), omitting the primary beam term $\Omega_{pp}$ in both cases.  We then compute
the ratio of the unnormalized $P(k)$ measurements before/after fringe-rate filtering and compare 
this ratio to the change in effective
beam area associated with the chosen fringe-rate filter.  Dividing the signal loss associated with
fringe-rate filtering by the change in beam area, we obtain a ratio of $1.016\pm0.011$ (1$\sigma$), 
indicating that these ratios are in agreement 
within the sample noise of our simulation. This numerically legitimizes the approach advocated above, where
we used approximate analytic arguments to argue that Equation \eqref{eq:P14est} can be used for power spectrum estimation provided the effective beam
squared integral is used instead of the beam squared integral for $\Omega_{pp}$.

\subsection{Minimizing polarization leakage}
\label{sec:polbeams}
\def\VXX{{V_{\rm XX}}}
\def\VYY{{V_{\rm YY}}}
\def\VI{{V_{\rm I}}}
\def\VQ{{V_{\rm Q}}}

\begin{figure*}\centering
\includegraphics[width=1.9\columnwidth]{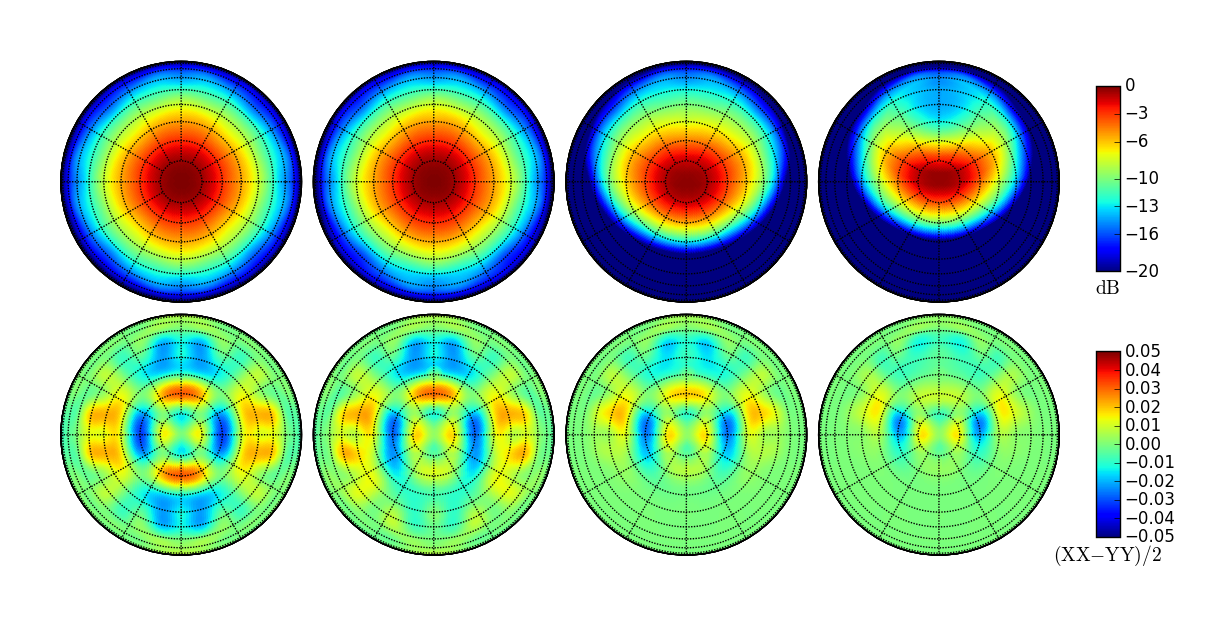}
\caption{
Beam response patterns illustrating the application of fringe-rate filtering to minimizing polarization leakage.
Panels from left to right illustrate the
response before matching XX and YY polarization beams, the response after such matching, 
the subsequent application of an optimal power-spectrum sensitivity filter, and the
application of a filter optimizing both sensitivity and polarization match.
The top row depicts the Stokes I beam response in logarithmic units; the
bottom row shows the polarization match ${\rm (XX-YY)}/2$ in linear units.
The fringe-rate filters optimizing sensitivity and polarization match correspond
to the red and green curves in Figure \ref{fig:fringe_rate_cut}, respectively.
The third column most closely corresponds to the fringe-rate filter applied in \citet{ali_et_al2015}.
}\label{fig:pol_match}
\end{figure*}

In the previous section, we maximized our sensitivity to the power spectrum under the assumption that the measurements were limited by instrumental noise. In practice, however, there may be other sources of noise or systematics that limit our constraints. One example of this is the cross-contamination between Stokes terms in interferometric polarization measurements. Minimizing such contamination is of importance for $21\,\textrm{cm}$ cosmology experiments that rely on
the spectral axis to probe the line-of-sight direction at cosmological distances.  For these
experiments, Faraday rotation combines
with a spurious coupling between Stokes terms (typically $Q$ to $I$) to produce polarization leakage whose 
spectral structure poses a worrisome foreground
to the cosmological signal \citep{jelic_et_al2008,jelic_et_al2010,jelic_et_al2014,bernardi_et_al2013,moore_et_al2013,moore_et_al2015}.  Current interferometers
targeting the 21cm signal at cosmological distances (LOFAR, MWA, PAPER, HERA, CHIME, LEDA) all employ linearly
polarized feeds, primarily because of their ease of construction and ability to co-locate elements sensitive to
orthogonal polarizations.  However, orthogonal linearly polarized feeds in practice have primary beam responses
that do not match.  As described in \citet{moore_et_al2013} and \citet{jelic_et_al2010}, if left uncorrected, the unmatched beam response 
between visibilities $\VXX$ and $\VYY$ measuring the XX and YY polarization products, respectively, is the 
dominant source of polarization leakage in the Stokes I measurement $\VI\equiv(\VXX+\VYY)/2$ for
linearly polarized feeds.

With an accurate beam model, it is trivial to rescale $\VXX$ and $\VYY$ 
so that the XX and YY beam responses match in one particular direction (typically the zenith).  Their sum, $\VI$, then
represents a perfect probe of the Stokes I parameter in that chosen direction, but will contain contamination
from $\VQ\equiv(\VXX-\VYY)/2$ in directions where the XX and YY beam responses do not match.
More precisely, suppose the XX polarization product has an antenna power beam of $A_{\rm XX}$, while the YY 
polarization product has an antenna power beam of $A_{\rm YY}$. The primary beam that we have denoted $A(\rhat)$ 
in previous sections is given by $(A_{\rm XX} + A_{\rm YY})/2$, which will be labeled as $A^I(\rhat)$ in this section to emphasize
its meaning as the antenna response to the Stokes I sky. Defining $A^Q(\rhat) \equiv (A_{\rm XX} - A_{\rm YY})/2$, the
Stokes I visibility takes the form (see, e.g., \citealt{moore_et_al2015})
\begin{eqnarray}
V_{b\nu}(t)= &&\int d\Omega \, {I_\nu(\rhat) A^I_\nu(\rhat,t) \exp \left[-i2\pi \frac{\nu}{c}  \mathbf{b}(t) \cdot \rhat\right]} + \nonumber \\
&& \int d\Omega \, {Q_\nu(\rhat) A^Q_\nu(\rhat,t) \exp \left[-i2\pi \frac{\nu}{c}  \mathbf{b}(t) \cdot \rhat\right]}, \qquad
\end{eqnarray}
where $Q_\nu$ denotes the Stokes Q sky at frequency $\nu$. The second term represents the Q to I leakage
in the visibility measurement, which does not vanish if $ A_{\rm XX} \neq A_{\rm YY}$, i.e., if there are
any asymmetries in the beams. For a single baseline at a given instant in time, this leakage term
cannot be eliminated. The heart of the problem is the impossibility of creating a match between a pair of two-dimensional functions (the
XX and YY beam responses) with a single degree of freedom (the amplitude of $\VXX$ relative to $\VYY$).  In order
to improve the match between polarization beams in interferometric measurements, many interferometric measurements
from distinct points in the $uv$ plane will have to be combined with appropriate weights to effect a re-weighting
of the sky along two dimensions.

A commonly used technique for correcting the mismatch between the XX and YY polarization beams is to separately
image these polarization products, correct each pixel in each image using modeled beam responses,
and then to sum the corrected images together to form a Stokes I map 
(e.g., \citealt{sullivan_et_al2012,bernardi_et_al2013,asad_et_al2015}).  Mathematically, this technique is identical to convolving the sampled
$uv$ plane by the Fourier transform of the directionally-dependent correction applied in the image domain. For an ideal
array that samples the $uv$ plane at scales significantly finer than the aperture of a single element, this technique can
in principle perfectly correct mismatches between the XX and YY polarization beams.
In practice, the success that can be achieved with this technique depends strongly on an array's $uv$ sampling pattern.

Take, for example, the case of a sparsely sampled $uv$ plane where the spacing between $uv$ samples is much greater than
the aperture scale of a single element.  In this case, the beam correction described above 
convolves each $uv$ sample with a kernel whose size scales roughly as the size of the aperture of a
single element in wavelengths.  Since this kernel is much smaller than the spacing between $uv$ samples, 
each point in the convolved $uv$ plane is dominated by the product of a kernel weight and a single visibility measurement.
As such, for a chosen $uv$ coordinate, the level of leakage in the Stokes I $uv$ plane effectively reduces to
what can be achieved using a single number to rescale $\VXX$ and $\VYY$ before 
summing.

In image domain, the baseline fringe pattern has been
integrated across the sky weighted by two different (XX, YY) beam responses to produce
a visibility.  This means
that an interferometric baseline for these two polarizations samples the $uv$
plane convolved by two different kernels --- the Fourier transforms of the XX,
YY beams, respectively.  Because the resulting $\VXX$ and $\VYY$ visibilities no
longer retain any direction-dependent information (they were integrated over
angle) $\VXX$ and $\VYY$ cannot be re-weighted in a way that produces pure Stokes I
response in all directions, except in the trivial case where the YY beam differs
from the XX beam by a direction-independent multiplicative factor.  Said
another way, given a single degree of freedom in the relative weighting of $\VXX$
and $\VYY$, it is not possible to guarantee a pure Stokes I response in more than
one direction.

As mentioned above, it is common practice in synthesis imaging 
to grid visibilities in the $uv$ plane
with convolving kernels that contain direction-dependent information.  Examples
include W-projection kernels \citep{cornwell_et_al2003}, the Fourier transform of the primary beam used in
optimal map making \citep{morales_matejek2009,bhatnagar_et_al2008}, and (relevant here) the 
Fourier transform of the inverse of
direction-dependent Mueller matrices \citep{tasse_et_al2013}.  Indeed, in the case of a completely
sampled $uv$ plane, the inverse Mueller beam kernel would perfectly match the XX
and YY beams.  However, as the sampling of the $uv$ plane becomes sparser, the
removal of samples removes degrees of freedom in weighting as a function of
direction.  In the limiting case of a very sparse array where the convolving
kernels of $uv$ samples do not overlap, we revert to the single
direction-independent weighting factor above.

Another way of thinking about this is to realize that, by gridding visibilities
in the $uv$ plane with a convolving kernel, we are attempting to reassign sky
intensity from the visibility, which was integrated over in equation \ref{eq:originalVis}, back to
its original location so that it may be re-weighted to match the XX and YY
polarization responses.  This inversion is only well posed in the case of a
completely sampled $uv$ plane.  As we lose samples, we lose the ability to
reconstruct the original sky intensity, and hence, to perfectly match the XX
and YY polarization responses as a function of direction.  In the case of a
sparsely sampled $uv$ plane, it is only possible to construct a pure Stokes I
response as a function of direction if one has sufficient additional
information about the sky (for example, if it sparsely populated by point
sources) that one can reconstruct the missing modes and hence, fully invert the
$uv$ plane.

For cases where $uv$ sampling falls somewhere between the sparse and the oversampled cases described above, the level
of primary beam correction that can be realized is more complicated.  Ultimately, the Fourier relationship between
the $uv$ plane and the image dictates that samples that are nearby to one
another in the $uv$ plane enable primary beam corrections on the largest angular scales, while samples that are farther
apart contribute to corrections on finer angular scales, with the orientation of the samples relative to one another
dictating the axis along which such corrections take effect in image domain.  Typically, earth-rotation synthesis
is required to sample the $uv$ plane densely enough to allow for effective beam correction for
diffuse structure\footnote{
As mentioned earlier in this section, it is possible to correct for direction-dependent polarization beams toward
sparse point sources using more widely separated visibility samples.  It is diffuse structure (which
is more localized in the $uv$ plane) that poses the greatest challenge.}
configurations are not dense enough to fully correct the beam even then.  One particularly relevant case that falls
in this last category are many of the maximum redundancy configurations currently favored by several $21\,\textrm{cm}$ cosmology experiments
for their sensitivity benefits \citep{parsons_et_al2012a,P14}.

However, even in the single-baseline case, earth-rotation synthesis provides dense $uv$ sampling along one direction: the
direction the baseline traverses in the $uv$ plane.  The appropriate
convolution kernel can combine samples along this track so as to correct the primary beam mismatch along one axis.
Of course, what we have just described---a convolution kernel acting along a time series
of samples from a single baseline---is precisely fringe-rate filtering.  Put another way, since fringe-rate filtering has the effect of modifying one's primary beams, it is possible to tailor fringe-rates to improve the match between the XX and YY polarization beams. In the case of sparse array sampling, fringe-rate filtering reproduces what can be achieved by independently
imaging the polarization products.  While this is not as effective at mitigating polarization leakage as can be achieved
through imaging in the dense sampling case, we will now show that it nonetheless represents a substantial improvement
over the naive summing of XX and YY visibility measurements.

Suppose for every frequency channel in our observations, we compute the RMS $A_{\rm XX} (\rhat)$ along
the spatial ring corresponding to each fringe-rate bin, using the relation $f \approx \rhat \cdot \mathbf{b}
\times \boldsymbol \omega_\Earth \nu / c$ derived in Section \ref{sec:overview}. The result is
a one-dimensional beam profile in fringe-rate space. Repeating the same exercise for $A_{\rm YY}$,
one can then form the ratio between the two profiles, quantifying the mismatch between the XX and YY
beams in a one-dimensional projection. This ratio is plotted for PAPER as the blue dashed curve in the
top panel of Figure \ref{fig:fringe_rate_cut}. If one then uses this curve as a set of fringe-rate weights
for the $\VYY$, the resulting effective beams will be more closely matched to one another, which we
will demonstrate later in this section when we quantitatively estimate polarization leakage for various fringe-rate filtering
schemes. Note that
we use the RMS rather than the beam itself because ultimately we seek a measurement
of the power spectrum, which is a squared quantity.

In the second column of Figure \ref{fig:pol_match}, we show the effective beam (top row) and the beam
mismatch (i.e., $A^Q (\rhat)$; bottom row) after performing our beam matching procedure. One sees
that the mismatch is slightly mitigated compared to the original unweighted case (leftmost column).
However, substantial power remains in the mismatched beam $A^Q (\rhat)$ because fringe-rate
filtering can only alter beam shapes in a one-dimensional family of fringe-rate rings (as we discussed in
Section \ref{sec:overview}). A perfect matching would require a set of two-dimensional weights.

Interestingly, the beam matching can be improved with one further weighting that 
is not specifically targeted at polarization matching, namely the sensitivity-optimized weighting
proposed in Section \ref{sec:PspecOptimization}. These were the weights applied in \citet{ali_et_al2015},
and the results of applying these weights to combined $V_I$ visibilities following the beam-matching procedure are shown
in the third column of Figure \ref{fig:pol_match}. While this extra weighting step was not designed
to deal with polarization in mind, it is still effective at mitigating leakage because it
downweights portions of the sky that have some of the worst beam mismatches.

Finally, one may use knowledge of the mismatched beam $A^Q (\rhat)$ to further inform one's weighting.
In particular, one may take the scheme that we have outlined so far (applying fringe-rate filters to first match XX and YY beams, then to maximize sensitivity), and add one final filter that downweights
portions of the sky where the mismatched beam is known to be large. In the fourth column of
Figure \ref{fig:pol_match}, we additionally weight each fringe-rate by the ratio of the RMS $A^I (\rhat)$ fringe-rate beam profile to the RMS $A^Q (\rhat)$ fringe-rate beam profile. The effective fringe-rate
weights that are applied to the Stokes I visibilities are given by the green curve of Figure \ref{fig:fringe_rate_cut}. From Figure \ref{fig:pol_match}, we see that this hybrid sensitivity- and leakage-
optimized weighting significantly reduces the beam mismatch.


In minimizing polarization leakage, one must be careful not to enact an overly aggressive
scheme that narrows the effective beam to such an extent that there is a substantial loss of
power spectrum sensitivity. To quantify this trade-off, consider a generalization of Equation
\eqref{eq:visCrossCorr} to include polarization leakage. Performing a derivation similar to the one
in Section \ref{sec:PspecOptimization} yields
\begin{equation}
\label{eq:LeakyCorrelFct}
\langle \widetilde{V}_i(\mathbf{u},\eta) \widetilde{V}_j(\mathbf{u},\eta)^*\rangle \approx S \left[P_I (\mathbf{k}) \Omega_{ij}^{II} + P_Q (\mathbf{k}) \Omega_{ij}^{QQ}\right],
\end{equation}
where $S \equiv ( B / X^2 Y) ( 2 k_B / \lambda^2 )^2$ and $(X k_x, X k_y, Y k_z) \equiv 2 \pi (u , v, \eta)$, as before. The indices $i$ and $j$ again index fringe-rate
bins. Beam integrals $\Omega^{QQ}$ and $\Omega^{II}$ are defined analogously to Equation \eqref{eq:Omega_ij_def},
but with superscripts indicating the Stokes components. The Stokes I power spectrum $P_I$ is the same as the
power spectrum $P$ defined in Section \ref{sec:PspecOptimization}, with the new notation serving only to
distinguish it from the Stokes Q power $P_Q$. In deriving Equation \eqref{eq:LeakyCorrelFct}, we made two
key assumptions. The first is that the I and Q contributions to the sky are on average uncorrelated. Empirically, 
this appears to be the case \citep{wieringa_et_al1993,gaensler_et_al2001,bernardi_et_al2003}, and physically, one expects this to be so since foreground interstellar medium clouds affect polarized and unpolarized
emission differently.
The second assumption is that the polarized sky is describable by a power spectrum. At some level, one expects
this assumption to fail, as there exist bright polarized sources that are not accounted for in a random realization
of some power spectrum. However, since our ultimate goal is to measure power spectra, it is reasonable to
define an effective Stokes Q power spectrum.

\begin{table*}[htbp]
   \caption{Sensitivity and polarization leakage metrics for different fringe-rate filters.}
   \centering
   \begin{tabular}{@{} rccccc @{}} 
      \toprule
       & Integrated & Square-integrated & Effective  & Polarization leakage& Normalized power\\
       & power beam [sr] & power beam $\Omega^{II}_\textrm{eff}$ [sr] &integration time [s] &  ($\Omega^{QQ}_\textrm{eff}/\Omega^{II}_\textrm{eff}$) & spectrum sensitivity\\
      \midrule
       No fringe-rate filter & 0.74& 0.32 & 420 & $1.64\times10^{-3}$  & 1. \\
       Polarization-matched fringe-rate filter & 0.74  & 0.32  & 420& $1.26\times10^{-3}$ & 1. \\
       Sensitivity-optimized fringe-rate filter & 0.51  & 0.24  & 2930 & $0.88\times10^{-3}$ & 1.9\\ 
       Polarization and sensitivity-optimized & 0.36  & 0.15  & 5570 & $0.68\times10^{-3}$ & 1.7 \\
      \bottomrule
   \end{tabular}

   \label{tbl:beam_metrics}
\end{table*}

Like before, rather than considering individual correlations between different fringe-rate bins, one may
insert fringe-rate filtered delay-space visibilities $\widetilde{V}^\textrm{filt} $ on the left-hand side of Equation \eqref{eq:LeakyCorrelFct}. The
same equation then holds with the fringe-rate bin indices omitted (since the fringe-rates have already
been combined in a weighted combination) and $\Omega^{QQ}$ and $\Omega^{II}$ replaced
by the square integral of the \emph{effective} beams, which we denote $\Omega^{QQ}_\textrm{eff}$ and $\Omega^{II}_\textrm{eff}$, respectively. One sees then that a suitable estimator of
the Stokes I power spectrum is
\begin{equation}
\widehat{P}_I (\mathbf{k}) =\frac{ | \widetilde{V}^\textrm{filt} (\mathbf{u}, \eta) |^2}{S \Omega^{II}_\textrm{eff}}.
\end{equation}
Ensemble-averaging both sides and inserting the fringe-rate filtered version of Equation \eqref{eq:LeakyCorrelFct} yields
\begin{equation}
\langle \widehat{P}_I (\mathbf{k}) \rangle = P_I (\mathbf{k}) + \frac{\Omega^{QQ}_\textrm{eff}}{\Omega^{II}_\textrm{eff}} P_Q (\mathbf{k}).
\end{equation}
The second term is the polarization leakage in our final power spectrum estimate. The key quantity is $\Omega^{QQ}_\textrm{eff}/\Omega^{II}_\textrm{eff}$, which quantifies the effectiveness of
one's polarization leakage suppression scheme. On the other hand, if the fringe-rate filtered
visibilities have an instrumental noise variance of $\sigma^2_\textrm{filt}$ (as we assumed in the previous section),
the instrumental noise errors $\Delta \widehat{P}_I$ in our power spectrum estimator are
\begin{equation}
\Delta \widehat{P}_I (\mathbf{k}) \equiv \left(\textrm{Var} \left[\widehat{P}_I (\mathbf{k}) \right]\right)^\frac{1}{2} = \frac{\sigma^2_\textrm{filt}}{S\Omega^{II}_\textrm{eff}}.
\end{equation}
For instrumental noise sensitivity, then, the crucial quantity is $\sigma^2_\textrm{filt} / \Omega^{II}_\textrm{eff}$. In Table \ref{tbl:beam_metrics} we list the value of this metric
(rightmost column) for the various weighting schemes considered in this section, normalized to the value
obtained with no fringe-rate filtering. We also show the integrated power beam $\int A^\textrm{eff} (\rhat) d\Omega$, the
square-integrated power beam $\Omega^{II}_\textrm{eff} \equiv \int A^\textrm{eff} (\rhat)^2 d\Omega$, the
effective (noise equivalent) integration time implied by each fringe-rate filter, and the fractional power spectrum
leakage $\Omega^{QQ}_\textrm{eff}/\Omega^{II}_\textrm{eff}$.

At this point, we should emphasize that because fringe-rate filtering performs
a weighted combination of samples over a wide time interval, the instrumental
noise errors $\Delta\widehat{P}_I$ are substantially correlated between nearby
times samples.  The number of independent modes preserved for power
spectrum analysis corresponds to the number and weighting of the filtered
fringe-rate modes.  Even without filtering,
quantifying the number of sampled modes is non-trivial, as the
time dependence of a baselines projection (and hence, $uv$ mode sampling) toward
a patch of sky varies substantially over the wide fields of view of
low-frequency interferometers.  As discussed in \citet{P14}, to
accurately account for the statistical interdependence of these modes in either the
fringe-rate filtering or square time-domain integration case, it is necessary
to use bootstrapping to determine errorbars. 

With only a polarization-matching filter applied weighting the YY polarization beam
to best match the XX polarization beam (second row of Table \ref{tbl:beam_metrics}; second column of Figure \ref{fig:pol_match}), we see that
the fractional polarization leakage is reduced from $1.64 \times 10^{-3}$ to $1.26 \times 10^{-3}$. Applying
an additional sensitivity weighting step from Section \ref{sec:PspecOptimization} (third row of Table \ref{tbl:beam_metrics}; third column of Figure \ref{fig:pol_match}) further reduces the fractional
polarization leakage to $0.88 \times 10^{-3}$, and though there is a 25\% reduction in effective
beam area, the power spectrum sensitivity is boosted by almost a factor of two. This improvement takes
into account the improvement in sensitivity of each independent power-spectrum sample, but
also the corresponding decrease in the number of independent time samples available. (Recall that fringe-rate filtering
effectively averages together nearby time samples, and thus returns a time series where instrumental noise
is no longer independent from time slice to time slice). Finally, applying
another filter to minimize leakage (fourth row of Table \ref{tbl:beam_metrics}; fourth column of Figure \ref{fig:pol_match}; green curve of Figure \ref{fig:fringe_rate_cut}), results in another $\sim25$\% decrease in polarization leakage with only a $\sim$10\% degradation
in sensitivity.  This degradation is associated with the 
longer integration times required to produce the narrow fringe-rate weighting profile needed for polarization
matching. This results in sensitive regions of the beam being underweighted because of their leakage.

Our final, best-performing weighting scheme for minimizing polarization was in some sense a rather
arbitrary weighting. This is unavoidable, since the power spectrum of low-frequency polarized emission is unknown at fine angular scales. We now show how our final weighting step may be optimized if the
polarized power spectrum is known. Suppose we estimate the power spectrum by forming weighted averages of the visibility cross-correlations between all
pairs of fringe-rate bins. The general form for such an estimator is given by Equation \eqref{eq:wVV}. However,
as we found in Section \ref{sec:PspecOptimization}, one may approximate the different fringe-rate bins as being
uncorrelated. The same approximation can be made here, allowing us to define $\mu^I$ and $\mu^Q$ by
letting $\Omega_{ij}^{II} \equiv \delta_{ij} \mu_i^I$ and $\Omega_{ij}^{QQ} \equiv \delta_{ij} \mu_i^Q$. With no
correlations between fringe-rate bins, it becomes sufficient to once again use Equation \eqref{eq:CountlessPhats}.
Taking the ensemble average of Equation \eqref{eq:CountlessPhats} but this time including polarization leakage
terms, i.e., inserting Equation \eqref{eq:LeakyCorrelFct} rather than Equation \eqref{eq:visCrossCorr} gives
\begin{equation}
\langle \widehat{P} \rangle = \left( S \sum_i w_i^2 \mu^I_i \right) P_I + \left( S \sum_i w_i^2 \mu^Q_i \right) P_Q.
\end{equation}
If, as before, we require our weights to be normalized so that $S \sum_i w_i^2 \mu^I_i =1$, the first term gives an
unbiased estimator of the Stokes I power spectrum. The second term is the bias in our final power spectrum estimate
due to leakage from Q to I. Ideally, the weights are chosen to mitigate this contribution. However, in attempting to
minimize this systematic, one must be careful not to pick weights that dramatically amplify the instrumental noise
contribution. We therefore choose to minimize an overall ``variance" that is the noise variance of Equation
\eqref{eq:NoiseVar} plus the square of the Stokes Q power spectrum bias (essentially adding the error bars
from instrumental noise and polarization leakage in quadrature). In other words, we minimize
\begin{equation}
\mathcal L = \sigma^4 \mathbf{m} \cdot \mathbf{m} + \left(S P_Q \right)^2 \left(\mathbf{m} \cdot \boldsymbol \mu^Q \right)^2 - \lambda \mathbf{m} \cdot \boldsymbol \mu^I,
\end{equation}
where we have introduced a Lagrange multiplier $\lambda$ to enforce our normalization constraint, and have grouped
the different fringe-rate weights into a vector $\mathbf{m}\equiv( w_1^2, w_2^2, \dots )$, with the beam overlap integrals similarly grouped into
vectors $\boldsymbol \mu^I$ and $\boldsymbol \mu^Q$. Now, notice that if we define
\begin{equation}
\mathbf{H} \equiv \sigma^4 \mathbf{I} + \left(S P_Q \right)^2 \boldsymbol \mu^Q (\boldsymbol \mu^Q)^t,
\end{equation}
our expression can be written as
\begin{equation}
\mathcal L = \mathbf{m}^t \mathbf{H} \mathbf{m} - \lambda \mathbf{m}^t \boldsymbol \mu^I.
\end{equation}
Differentiating this, setting the result to zero, and solving for the normalized weights gives
\begin{equation}
\label{eq:PolOptWeights}
\mathbf{m} =  \frac{\mathbf{H}^{-1}  \boldsymbol \mu^I}{ S (\boldsymbol \mu^I)^t  \mathbf{H}^{-1}  \boldsymbol \mu^I}.
\end{equation}
This expression involves $\mathbf{H}^{-1}$, and can be evaluated explicitly in our case using the
Woodbury formula, giving
\begin{equation}
\label{eq:Hinv}
\mathbf{H}^{-1} = \sigma^{-4} \left[ \mathbf{I} - \frac{P_Q^2\boldsymbol \mu^Q (\boldsymbol \mu^Q)^t }{P_N^2 + P_Q^2 \boldsymbol \mu^Q \cdot \boldsymbol \mu^Q} \right],
\end{equation}
where $P_N$ is the noise power spectrum, which is equal to the RMS visibility
noise divided by by $S$. If the sky were completely unpolarized ($P_Q = 0$) or there were no polarization leakage due to mismatched beams ($\boldsymbol \mu^Q = 0$), then $\mathbf{H}^{-1} \propto \mathbf{I}$, and the optimal weighting would take the form
$\mathbf{m} \propto \boldsymbol \mu^I$. Recalling that each component of a $\boldsymbol \mu$ vector corresponds
to a different fringe rate, we see that $\mathbf{m} \propto \boldsymbol \mu^I$ is precisely the result derived in Section
\ref{sec:PspecOptimization}, where each fringe-rate bin was weighted by the integrated square of the beam within the bin.

With polarization leakage, Equation \eqref{eq:PolOptWeights} shows that as a first step, one should still weight by
the square-integrated beam within each fringe-rate bin. However, one should then further weight by the more complicated
form for $\mathbf{H}^{-1}$ given by Equation \eqref{eq:Hinv}. To gain some intuition for
this operation, consider the limit of zero instrumental noise. One then obtains 
\begin{equation}
\mathbf{H}^{-1} \Big{|}_{P_N =0} \propto \mathbf{I} - \boldsymbol \mu^Q (\boldsymbol \mu^Q \cdot \boldsymbol
\mu^Q )^{-1} (\boldsymbol \mu^Q )^t,
\end{equation}
which we immediately recognize as a projection matrix that projects out the vector $\boldsymbol \mu^Q$. Since the
components of $\boldsymbol \mu^Q$ encode the polarization leakage response in various
fringe-rate bins, $\mathbf{H}^{-1}$ projects out linear combinations of
fringe rates that are indicative of Q to I leakage due to beam asymmetries. At the other
extreme, if the polarization power spectrum is weak compared to the noise power spectrum,
we have
\begin{equation}
\mathbf{H}^{-1} \Big{|}_{P_N \gg P_Q} \propto \mathbf{I} - \left(\frac{P_Q}{P_N} \right)^2 \boldsymbol \mu^Q  (\boldsymbol \mu^Q )^t,
\end{equation}
which is similar to a projection operator, but with the unwanted or projected piece tempered
by the square of the small parameter $P_Q / P_N$. Intuitively, even modes that are
contaminated by polarization leakage contain cosmological information, and it is advantageous
to avoid too drastic a projection if possible. An overly aggressive subtraction of leakage
modes results in a loss of cosmological signal, which when corrected for by the altered
normalization of the power spectrum, results in a magnification of the error bars. For the
general case of intermediate noise levels, the $\mathbf{H}^{-1}$ matrix smoothly interpolates
between the two extremes.

While mathematically optimal, the fringe-rate weighting proposed here is currently difficult
to put into practice. This is because the $Q$ power spectrum $P_Q$ has yet to be positively
measured at low frequencies. For example, recent measurements in \citet{moore_et_al2015}
provide only upper limits on $P_Q$, at least at the $k$-scales that are the
most promising for a first detection of the 21cm signal.
%
%

\subsection{Minimizing Instrumental Systematics and Off-Axis Foregrounds}
\label{sec:foregrounds}

\begin{figure}\centering
\includegraphics[width=.9\columnwidth]{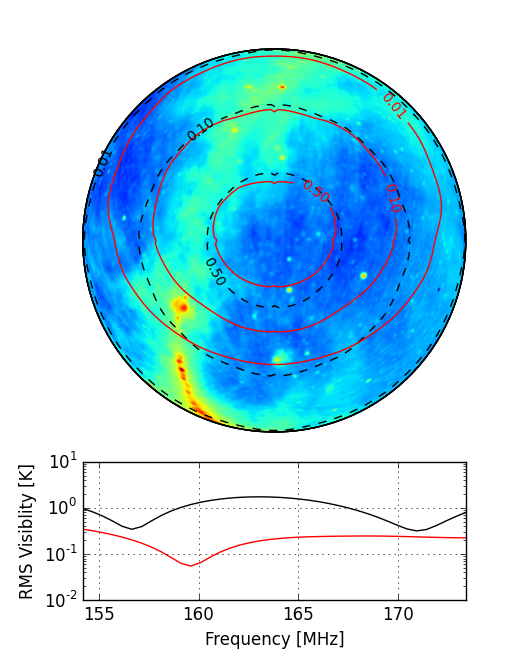}
\caption{
Top: A projection of the global sky model of \citet{deoliveiracosta_et_al2008} as viewed from
latitude $-30^\circ$ at a local sidereal time of 05:44, overlaid with the 
PAPER beam response contours before (black) and after (red) the application of a fringe-rate filter
optimized for sensitivity.  
Bottom: a simulation of the corresponding RMS visibilities for a 30-m east-west baseline before (black) and
after (red) the same fringe-rate filtering.
The reduced presence of foregrounds in the simulated visibilities after fringe-rate filtering
can be understood as coming from the spatial filter enacted by fringe-rate filtering.
}\label{fig:gsm}
\end{figure}

A final application of beam sculpting with fringe-rate filters targets the suppression of systematics in data.
We will consider two systematics: additive phase terms associated with instrumental crosstalk, and sidelobes
associated with celestial emission outside of the primary field of interest.  Both of these applications are 
closely aligned with the original application of fringe-rate filters described in \citet{parsons_backer2009}. Reduction of sidelobe contamination (as well as radio frequency
interference mitigation) using a similar technique was also discussed in \citet{offringa_et_al2012}.

For the purposes of this discussion, we consider instrumental crosstalk to be a spurious 
correlation introduced between otherwise uncorrelated signals
as a result of electromagnetic coupling in the instrument (typically between adjacent, unshielded signal lines)
or because a non-celestial source has injected a correlated signal (e.g. switching noise on power supplies).
Although crosstalk can be suppressed using phase switching \citep{ryle1952}, it is always present at some level
in interferometric observations.  If it is temporally stable, however, it is possible to significantly
suppress crosstalk in data by averaging visibilities over a long period (so that the fringing celestial
signal washes out) and then subtracting the average complex additive offset from the data.  This 
technique has
long been applied to, e.g., PAPER observations \citep{parsons_et_al2010,pober_et_al2013,J14,P14,ali_et_al2015}.

As a time-domain filter, this crosstalk removal technique can also naturally be understood as a notch filter
for removing signals with zero fringe-rate.  Because crosstalk removal uses a finite time interval for computing
the average, applying this notch fringe-rate filter has the effect of removing emission from the region of
sky corresponding to the zero fringe-rate bin.  As illustrated in Figure \ref{fig:fringe_contours}, for our
fiducial baseline,
this corresponds to the unshaded region intersecting the south celestial pole.
For PAPER, this region is sufficiently low in the beam that its removal has little impact, but in general, 
subsequent analysis of crosstalk-removed data may require accounting for the beam-sculpting effects of
the crosstalk removal filter.

Thus, when considering instrumental systematics, there may be additional criteria that influence one's
choice of fringe-rate filter besides optimizing signal-to-noise; one may choose to excise the zero fringe-rate
bin to improve data quality at a very modest cost to sensitivity.  Similarly, it is common to encounter situations
where celestial emission that is low in the primary beam is bright enough to introduce undesirable 
sidelobe structure or other systematics in observations targeting an area nearer to beam center.  In this case,
one may again find it desirable to depart from optimal weighting derived in Section \ref{sec:PspecOptimization}
by further down-weighting regions of low sensitivity in order to gain improvements in foreground systematics.
This application of fringe-rate filtering is particularly relevant for 21cm cosmology experiments where approximately
Gaussian signals are overlaid with highly non-Gaussian foregrounds.  Fringe-rate filters that are informed by 
the angular structure in foreground models can substantially suppress foreground systematics while having little
impact on a statistically isotropic Gaussian signal.

In Figure \ref{fig:gsm} we illustrate this technique using simulations of a single baseline. The sky model
that we employ is that of \citet{deoliveiracosta_et_al2008}, and in the top panel of the figure we show
the field of view of our simulated interferometer at a particular instant in time. Primary beam contours
for the PAPER instrument are shown in dashed black, and one sees that there is substantial power from
bright Galactic plane emission. In solid red are the contours for the effective primary beam following
the application of a power-spectrum optimized fringe-rate filter (see Section \ref{sec:PspecOptimization}).
The reduced effective beam size means that less of the Galactic plane contributes to the measured
power. This can be seen in the bottom panel of Figure \ref{fig:gsm}, where we show the corresponding
RMS visibilities. The red curve is clearly reduced compared to the black curve, indicating
reduced foreground contamination. If desired, the contamination can be further mitigated by picking
a different set of fringe-rate weights that downweight the edges of the primary beam even more. However,
this extra foreground suppression must be carefully balanced with the increase in instrumental noise error
bars that inevitably results from the narrowing of the effective beam beyond the size implied by the optimized procedure of
Section \ref{sec:PspecOptimization}.




\section{Fringe-rate filtering as mapmaking from time-ordered data}
\label{sec:Mapmaking}

In the previous sections, we have focused on applications of fringe-rate filtering that operate on a single baseline basis. These applications are particularly powerful for maximally redundant arrays such at PAPER, which have most of their sensitivity concentrated in multiple identical copies of a small handful of baseline types. By design, maximally redundant arrays are not optimized for imaging, which instead require arrays that sample a large number of unique baselines. In this section, we turn our attention to such arrays, tackling the imaging (i.e., mapmaking) problem for multi-baseline arrays. We will find once again that the fringe-rate space is particularly well-suited for implementing time integration for interferometric data.


Suppose our time-ordered visibilities are grouped into a measurement vector
$\vis$ of length $N_b N_t$, where $N_b$ is the number of baselines, and $N_t$
is the number of snapshots taken in time.  If we represent the true sky as a
vector $\x$ of length $N_\textrm{pix}$, and our instrument's response as a
matrix $\A$ of size $N_b N_t \times N_\textrm{pix}$, the measurement equation
is given by
\begin{equation}
\label{measEqn}
\vis = \A \x + \mathbf{n},
\end{equation}
where $\mathbf{n}$ is a noise vector.  Note that in this general form, Equation
\eqref{measEqn} is not basis-specific.  For example, while it is often useful
to think of $\x$ as a vector containing a list of temperatures in a set of
pixels on the sky (hence the variable name $N_\textrm{pix}$), it is equally
valid to employ another basis, such as spherical harmonics.  Similarly, while
we call $\vis$ the time-ordered data, it need not be a time series, and in
fact, we will see that a description in fringe-rate space is in fact quite natural.

Given our measurement $\vis$, the optimal estimator $\xhat$ of the true sky
$\x$ is given by \citep{T97mapmaking,Morales2009,dillon_et_al2015}
\begin{equation}
\label{optEst}
\xhat = \mathbf{M} \A^\dagger \N^{-1} \vis,
\end{equation}
where $\mathbf{M}$ is some invertible matrix chosen by the data analyst, 
the dagger signifies an adjoint, and
$\N$ is the noise covariance matrix, defined as $\langle \mathbf{n}
\mathbf{n}^\dagger \rangle$, with angled brackets denoting an ensemble average.
Again, our vector/matrix expressions are basis-independent, so even though the
formation of $\xhat$ is often described as ``mapmaking", it need not correspond
to spatial imaging in the traditional sense of the word. A similar approach
can also be found in ``A-projection" algorithms \citep{bhatnagar_et_al2008,
bhatnagar_et_al2013,tasse_et_al2013}.

The error bars on the estimator $\xhat$ are obtained by computing the square root
of the diagonal elements of the covariance $\boldsymbol \Sigma$, which is given by
\begin{equation}
\label{eq:sigma}
\boldsymbol \Sigma \equiv \langle (\x - \xhat) ( \x - \xhat)^\dagger \rangle = \mathbf{M} \A^\dagger \N^{-1} \A\mathbf{M}^\dagger.
\end{equation}
With a suitable choice of $\mathbf{M}$, the estimator given by Equation \eqref{optEst}
minimizes the variance. Regardless of one's choice, however, Equation \eqref{optEst} 
can be shown to be lossless \citep{T97mapmaking}, in the sense that any quantities (such as power
spectra) formed further downstream in one's analysis will have identically
small error bars whether one forms these data products from $\xhat$ or chooses
to work with the larger and more cumbersome set of original data $\vis$.
%
%
%

In principle, Equation \eqref{optEst} is all that is needed to optimally
estimate the true sky.  One simply forms the relevant matrices and performs the
requisite matrix inversions and multiplications.  However, this is
computationally infeasible in practice, given that modern-day interferometers
are comprised of a large number of baselines operating over long integration
times, resulting in rather large matrices.  This is what motivated the authors
of \cite{Shaw2013} to propose their $m$-mode formalism, essentially rendering
many of the relevant matrices sparse, making them computationally easy to
manipulate.  While the $m$-mode formalism is a general framework that can be
used to solve a variety of problems (such as mitigating foreground
contamination), our goal here is to develop similarly convenient techniques for
the mapmaking problem (i.e., the formation of $\xhat$), with much detail
devoted to the intuition behind how our optimal estimator operates for an
interferometer.

\subsection{The general sub-optimality of time integration}
\label{timeSubOpt}

We begin by showing that it is suboptimal to make maps by integrating visibilities in time.
Writing out Equation \eqref{eq:originalVis} for the visibility $V_{b\nu}(t)$ with an explicit coordinate system, we have\begin{eqnarray}
\label{generalVis}
V_{b\nu}(t) &&= \int A_\nu(\rhat, t) I_\nu(\rhat) \exp \left[ -i 2 \pi \left( \frac{b_y}{\lambda} \cos \eta \sin \delta \right) \right] \nonumber \\
&& \times  \exp \left[ -i 2 \pi \left( \frac{b_0}{\lambda} \cos \delta \sin(\alpha - \omega_\Earth t ) \right) \right]  d\Omega +n(t), \qquad
\end{eqnarray}
where $n(t)$ is the instrumental noise, $\alpha$ and $\delta$ are the right ascension and declination, respectively, $\eta$ is the geographic latitude of the
array, and $b_0 \equiv \sqrt{b_x^2 + b_y^2 \sin^2 \eta}$, where $b_x$ and $b_y$
are the east-west and north-south baseline lengths, respectively.  We have chosen
our definition of $t=0$ to conveniently absorb an arbitrary constant phase. Like before,
 we are assuming that the primary beam is fixed with
respect to local coordinates and translates azimuthally on the celestial
sphere.  We additionally assume that the baseline is phased to zenith.  In
other words, Equation \eqref{generalVis} describes an interferometer observing
in a drift-scan mode.

To see how integrating in time may be suboptimal, consider a simplified, purely
pedagogical thought experiment where our interferometer consists of a single
east-west baseline ($b_y=0$) situated at the equator ($\eta = 0$).  For the
primary beam, suppose we have a beam that is extremely narrow in the polar
direction, so that $A_\nu(\rhat, t) \equiv \delta^{D}(\delta)
A_\nu^\alpha(\alpha-\omega_\Earth t)$, where $ \delta^{D}$ signifies a Dirac delta
function.  Plugging these into restrictions into our
equation, we obtain
\begin{eqnarray}
\label{simpVis}
V_{b\nu} (t) && = \int  \, A_\nu^\alpha(\alpha-\omega_\Earth t)  I_\nu \left( \delta = 0, \alpha \right) \nonumber \\
&&\times \exp \left[ -i 2 \pi  \frac{b_x}{\lambda} \sin(\alpha - \omega_\Earth t ) \right] d\alpha+ n(t).
\end{eqnarray}
For a single baseline, the function $V_{b\nu} (t)$ is precisely the continuous version
of the discrete data vector $\vis$. To obtain $\vis$, then, one would simply sample
$V_{b\nu} (t)$ discretely in time. For a multi-baseline array, forming $\vis$ involves following
the above procedure for each baseline, and then concatenating the resulting vectors
to form a single long $\vis$ vector. To make our analytic manipulations more convenient,
however, we will keep $t$ a continuous variable, so that $\vis$ is a hybrid quantity,
discrete in baseline but continuous in time. Acting on $\vis$ by a matrix then involves
summing over baselines and integrating over time.

Identifying $n(t)$ and $I_\nu(\theta= \pi / 2, \varphi)$ as
the continuous versions of $\mathbf{n}$ and $\x$ respectively, the rest of
Equation \eqref{simpVis}'s integrand can be interpreted as the continuous version of $\A$.  We
can model the noise covariance between baselines $b$ and $b^\prime$, at times
$t$ and $t^\prime$ as
\begin{equation}
\label{eq:noiseCovar}
N_{bb^\prime} (t, t^\prime) = \sigma^2 \delta_{bb^\prime} \delta^D(t-t^\prime),
\end{equation}
where $\sigma$ is an RMS noise level assumed to be uncorrelated in
time and uncorrelated between baselines.

To see how the optimal prescription of Equation \eqref{optEst} combines
information from different times, we need only evaluate $\A^\dagger \N^{-1}
\vis$, for the $\mathbf{M}$ matrix has no time index, so its application has no impact
on how time-ordered data is combined. In our toy model, we
have
\begin{equation}
\label{toyComb}
\left( \A^\dagger \N^{-1} \vis \right)_{\alpha} \! = \! \sum_b \!\!  \int \frac{dt}{\sigma^2} \,  A_\nu^\alpha(\alpha-\omega_\Earth t) \, e^{ i 2 \pi  \frac{b_x}{\lambda} \sin(\alpha - \omega_\Earth t ) } V_{b\nu}(t),
\end{equation}
where the $\alpha$ variable serves as the continuous version of a discrete
vector index.  This expression shows that the optimal, minimum variance
prescription does not call for the integration of visibilities in time.
Instead, our expression calls for the \emph{convolution} of the visibility data
with a kernel that is specified by the primary beam shape and the baseline.

Now, recall from our application of the convolution theorem in previous sections
that for interferometric data, convolution in time is equivalent to multiplication 
in fringe-rate space.  Equation \eqref{toyComb}
therefore suggests that the optimal way to combine different time samples is to
express visibilities in fringe-rate space, and then to weight different
fringe-rates appropriately before summing.  We will develop this method for mapmaking more generally in Section \ref{fringeRateIntro}.


\subsection{The special case where integrating in time is optimal}

Before proceeding, it is instructive to establish the special case where time
integration is the optimal technique for an initial data reduction step in mapmaking, since it is frequently employed in the
literature.  An inspection of Equation \eqref{toyComb} shows that were it not
for the time-dependence in the primary beam and the time-dependence of the sky
moving through a baseline's fringes, the optimal recipe would indeed reduce to
an integration of visibilities in time.  Finding the limit where time
integration is optimal is then equivalent to finding a special case where the
aforementioned time-dependences vanish.

Recall that in our previous example, the primary beam had a time-dependence
only because our thought-experiment consisted of a drift-scan telescope, whose
measurement equation was written in coordinates fixed to the celestial sphere.
Instead of this, suppose one had a narrow primary beam that tracked a small
patch of the sky.  The primary beam would then have a fixed shape in celestial
coordinates, and $A_\nu(\rhat, t)$ would simply become $A_\nu(\rhat)$ in Equation
\eqref{generalVis}.  To attempt to nullify the time-dependence of fringes
sweeping across the celestial sphere, one may phase the visibilities in a
time-dependent way, essentially tracking the center of the patch as it moves
across the sky.  Putting this all together and assuming that the primary beam
is narrow enough to justify a flat-sky approximation, the measurement equation
becomes
\begin{eqnarray}
\label{flatSkyVis}
V_{b\nu}(t) &&= \int A_\nu(\rhat) I_\nu(\rhat) \exp \left[ -i 2 \pi \left( \frac{b_0}{\lambda}  \sin(\alpha - \omega_\Earth t )\right) \right] \nonumber\\
&&\!\!\! \times \exp \left[-i 2 \pi \left( \frac{b_y}{\lambda} \cos \eta \sin \delta \right) + i \psi(t) \right] d\Omega +n(t), \qquad
\end{eqnarray}
where we have assumed for simplicity that the center of our small field is
directly above the equator, and that a time-dependent phase $\psi(t)$ has been
applied.  With this, the optimal combination of time-ordered data becomes
\begin{eqnarray}
\left( \A^\dagger \N^{-1} \vis \right)_{(\delta,\alpha)}^\textrm{flat} &&  = \frac{A_\nu(\delta,\alpha)}{\sigma^2} e^{-i 2 \pi \frac{b_y}{\lambda} \cos \eta \sin \delta}  \nonumber  \\
&&\!\!\! \times \sum_b \int  dt \,\,e^{ -i 2 \pi  \frac{b_x}{\lambda} \sin(\alpha - \omega_\Earth t ) +i \psi (t) } V_{b\nu}(t). \qquad
\end{eqnarray}
This is still not quite a simple average in time because there is no choice of
$\psi(t)$ that can cancel out the time-dependence of $\sin (\alpha -
\omega_\Earth t)$ for all $\varphi$ and all $t$.  Another way to phrase the
problem is to note that even in the flat-sky approximation, one cannot expand
Taylor expand $\sin (\alpha - \omega_\Earth t)$ over long observation times.
With short observations, however, an expansion is justified. Performing this
expansion, invoking the narrow field-of-view approximation in the $\alpha$ direction, and picking
$\psi(t) = -2 \pi \frac{b_x}{\lambda} \omega_\Earth t$ gives
\begin{eqnarray}
\left( \A^\dagger \N^{-1} \vis \right)_{(\delta,\alpha)}^\textrm{flat,short} =  &&\frac{A_\nu(\delta,\alpha)}{\sigma^2} e^{-i 2 \pi \left( \frac{b_y}{\lambda} \cos \eta \cos \delta + \frac{b_x}{\lambda} \sin\alpha \right)} \nonumber \\
&& \times \sum_b   \int  dt \, V_{b\nu}(t),
\end{eqnarray}
which is a simple averaging in time.  In short, then, integrating in time is an
optimal way to combine time-ordered data if a number of criteria are met:
the flat-sky approximation must hold (which can be achieved, for example, by having a small primary beam), the primary beam must track the field,
the visibilities must be phased to track the center of the field, and the
observations must be short.

\subsection{Optimized fringe-rate filtering for imaging}
\label{fringeRateIntro}

We now proceed to derive the optimal prescription for combining time-ordered
data, which will naturally give rise to fringe-rate filtering.  Because we will
\emph{not} be invoking the same approximations as we did in previous sections,
we will begin with Equation \eqref{generalVis}.  From
our toy example (Equation \ref{toyComb}), we know that fringe-rate space (the
Fourier dual of time) is a promising space in which to combine time-ordered
data.  Here, we make the assumption that we are dealing with a full sidereal
day's worth of data, so that the visibility in fringe-rate space is given by
\begin{equation}
\label{eq:FringeTransformDef}
\overline{V}_{b\nu} (f) \equiv \frac{1}{T_\Earth} \int_{-T_\Earth / 2}^{T_\Earth /2} dt \exp \left( - 2 \pi i f t \right) V_{b\nu} (t),
\end{equation}
where $f$ is the fringe-rate, and $T_\Earth = 2 \pi / \omega_\Earth$ is the
Earth's rotation period. It is natural to work in fringe-rate bins such that
the $n$th bin is given by $f_n \equiv n / T_\Earth$, where $n$ is an
integer.  The measurement in the $n$th bin is then given by
\begin{eqnarray}
\label{fringeVis}
\overline{V}_{b\nu} (f_n) &&= \int d\Omega \, T(\rhat) e^{-i 2 \pi  \frac{b_y}{\lambda} \cos \eta \sin \delta} \nonumber \\
&& \!\!\!\times  \int_{-\frac{T_\Earth}{ 2}}^{\frac{T_\Earth}{2}} \frac{dt}{T_\Earth} \, B(\rhat, t) e^{ -i  \frac{2 \pi n t}{T_\Earth} +i  \frac{2\pi b_0}{\lambda} \cos \delta \sin \left( \omega_\Earth t - \alpha \right)}, \qquad
\end{eqnarray}
where we have temporarily omitted the additive noise term to avoid mathematical
clutter.  To proceed, we make some simplifying assumptions (although only some
of which are absolutely required).  First, assume that we are once again
considering a drift-scan instrument.  If the primary beam shape is
approximately separable, we can then say
\begin{equation}
\label{eq:SepBeam}
B(\rhat, t) \equiv B_\delta (\delta) B_\alpha (\alpha - \omega_\Earth t),
\end{equation}
where $B_\alpha$ is a function with period $2\pi$.  Taking advantage of this periodicity, we can write the beam as
\begin{equation}
B(\rhat, t) = B_\delta (\delta) \sum_q \overline{B}_q e^{-iq \alpha} e^{i q \omega_\Earth t},
\end{equation}
where $\overline{B}_q \equiv \int \frac{d\alpha}{2\pi} B_\alpha(\alpha)
e^{i q\alpha}$ is the $q^{th}$ Fourier coefficient.  Plugging this into
Equation \eqref{fringeVis} and making the substitution $\psi \equiv
\omega_\Earth t - \varphi$, one obtains
\begin{eqnarray}
\overline{V}_{b\nu} && (f_n)= \int d\Omega \, T(\rhat) B_\delta (\delta) e^{-i 2 \pi  \frac{b_y}{\lambda} \cos \eta \sin \delta} \nonumber \\
&& \times \sum_q \frac{\overline{B}_q e^{-i n \varphi}}{2 \pi} \int_{-\pi -\varphi}^{\pi+\varphi} d\psi \, e^{i (q-n) \psi +i \frac{2 \pi b_0}{\lambda} \cos \delta \sin \psi}. \qquad
\end{eqnarray}
Now, the integral over $\psi$ is of a periodic function over one
period.  We may therefore freely shift the limits of the integral by a constant
amount without affecting the result.  In particular, we may remove the
$+\varphi$ terms in the limits (the only restriction being that having performed a $\varphi$-dependent shift, it is no longer legal to permute the various integrals), and
the result is a standard integral form for a Bessel function $J$ of the first
kind:
\begin{eqnarray}
\label{fringeBessel}
\overline{V}_{b\nu} (f_n) =&& \int \frac{d\Omega}{2 \pi}\, T(\rhat) B_\delta (\delta) e^{-i 2 \pi  \frac{b_y}{\lambda} \cos \eta \sin \delta}
e^{-i n \alpha} \nonumber \\ 
&& \times \sum_q \overline{B}_q  J_{n-q} \left( \frac{2 \pi b_0}{\lambda} \cos \delta \right).
\end{eqnarray}
Several features are of note here.  For wide primary beams, $\overline{B}_q$
is sharply peaked around $q=0$, so the terms following the sum over $q$
essentially amount to $J_n ( 2 \pi b_0 \cos \delta / \lambda )$.  Now, notice
that the argument of the Bessel function is bounded, always lying between $\pm
2\pi b_0/ \lambda$.  For large $n$ (high fringe-rate bins), then, one can use
the small argument asymptotic form for $J_n$,\begin{equation}
J_n \left( \frac{2 \pi b_0 \cos \delta}{ \lambda} \right) \approx \frac{1}{n!} \left( \frac{ \pi b_0 \cos \delta}{ \lambda} \right)^n,
\end{equation}
which is a sharply decreasing function of $n$ for large $n$.  This means that
there must be very little sky signal at high fringe-rate bins, which is yet another
reflection of the salient feature that we have emphasized throughout this paper:
high fringe-rate bins constitute noise-dominated modes, since the Earth's rotation
rate works in conjunction with the baseline length to impose a maximum fringe-rate
for sources locked to the celestial sphere.

Putting together the optimal prescription as we did above, we have
%
\begin{eqnarray}
\label{eq:AdagNinvv}
\left( \A^\dagger \N^{-1} \vis \right)_{\delta,\alpha} && = \frac{B_\delta^* (\delta)  \cos \delta}{2\pi \sigma^2} \sum_{b,n} e^{i n \alpha} e^{i 2 \pi  \frac{b_y}{\lambda} \cos \eta \sin \delta} \nonumber \\
&& \times \sum_{q} \widetilde{B}_q^* J_{n-q} \left( \frac{2 \pi b_0}{\lambda} \cos \delta \right) \overline{V}_{b\nu} (f_n). \qquad
\end{eqnarray}
In words, this recipe instructs us to move into fringe-rate space (where the
sky emission is already concentrated in $f_n$) and to further downweight by
$\sum_{q} \widetilde{B}_q^* J_{n-q} \left( \frac{2 \pi b_0}{\lambda} \cos \delta \right)$, which, as we have argued above, is small for high fringe
rates.  Thus, filtering away the high fringe-rates is the optimal way to combine
time-ordered data from an interferometer.

In summary, we see that downweighting high fringe-rates once again appears as an integral part of an optimal
recipe, this time for mapmaking. While we made some assumptions (such as a separable beam) for analytic
convenience, our qualitative conclusions are general. To generalize our treatment, one can simply return
to Equation \eqref{fringeVis}, numerically evaluating the integral over time. One can then read off an expression
for $\mathbf{A}$ from the remaining integral, and proceed as before.

\section{Conclusion}
\label{sec:conclusion}

In this paper, we have revisited the concept of filtering the visibility time-series
measured by an interferometric baseline that was presented in \citet{parsons_backer2009}.
Using a mapping between the timescale of variation in visibility data and position
on the sky for a chosen baseline, we have seen that the rectangular time windows typically
used when integrating visibilities are almost always sub-optimal (particularly for wide-field, drift-scan instruments), and motivate 
filtering on the basis of fringe rate
as a step for optimally combining time-ordered visibility data.  In Section \ref{sec:Mapmaking}, we found
that fringe-rate filtering also naturally arises as part of optimized mapmaking prescriptions such as those
described in \citet{T97mapmaking}, \citet{morales_matejek2009}, or \citet{dillon_et_al2015}.

We also showed that fringe-rate filtering can alternately be interpreted as a per-baseline
operation for sculpting the primary beam along one dimension.
Using analytic
derivations and simulations, we highlight several important applications of such beam
sculpting.  One key application for 21cm cosmological experiments starved for sensitivity
is the ability to re-weight visibility data according to the signal-to-noise ratio in each fringe-rate bin.
As shown in Section \ref{sec:PspecOptimization}, pre-processing the visibilities in such a way
allows an optimal (minimum variance) estimate of the power spectrum to be formed by
squaring visibilities time-slice-by-time-slice before averaging the results together. This
somewhat surprising result allows a data analyst to deal directly with visibilities until
the final squaring step, allowing for cleaner diagnoses of instrumental systematics while
avoiding potential analysis-induced systematics associated with gridding data on the $uv$ plane.
%
Other important applications include improving the match between polarization beam to
reduce polarization leakage, and down-weighting areas low in the primary beam
to reduce systematics from off-axis foregrounds.

In \citet{ali_et_al2015}, the fringe-rate filtering techniques presented here are applied to
observations from the PAPER array as part of their power-spectrum analysis pipeline.  The
results highlight the power of fringe-rate filtering in 21cm cosmology applications.
Given its efficiency, flexibility, and close alignment with the natural observing
basis of radio interferometers, we anticipate that fringe-rate filtering may be
an important analysis tool for current 21cm experiments, as well as future instruments
such as the Hydrogen Epoch of Reionization Array (HERA; \citealt{pober_et_al2014}) and
the Square Kilometre Array (SKA; \citealt{carilli2014}).

\section{Acknowledgments}

It gives us great pleasure to thank James Aguirre, Josh Dillon, Gianni Bernardi, Daniel Jacobs, Saul Kohn, David Moore, 
Miguel Morales, and Jonathan Pober for helpful discussions.  
This research was supported in part by the NSF CAREER award No. 1352519 and the NSF AST grant No. 1129258.


\bibliographystyle{apj}
\bibliography{fringe_filter}

\end{document}